\pdfoutput=1
\documentclass[12pt]{article}
\usepackage[utf8]{inputenc}
\usepackage{cite}
\usepackage{hyperref}
\hypersetup{colorlinks=true,linkcolor=blue2,citecolor=red2,urlcolor=green2,pdfencoding=auto,linktocpage}
\usepackage{amsmath,amssymb,amsbsy,amstext,amsthm,simplewick,amsfonts}
\usepackage{mathrsfs}
\usepackage{graphicx}
\usepackage{wrapfig}
\usepackage{upgreek}
\usepackage{bm} 
\usepackage{framed}
\usepackage{bbm}
\usepackage{textcomp}
\usepackage{adjustbox}
\usepackage{makecell}
\usepackage{tcolorbox}
\usepackage{empheq}
\usepackage[normalem]{ulem}
\usepackage{enumitem}
\usepackage{braket} 
\usepackage{array}
\usepackage{dsfont}
\usepackage{physics}
\usepackage{ulem}
\usepackage{xcolor}
\usepackage{caption}
\usepackage{subcaption}
\usepackage{tocloft}
\usepackage{tikz}
\usepackage{slashed}
\usepackage{bbm}
\usepackage{lmodern}
\usepackage{booktabs} 
\usepackage{amsmath}  
\usepackage{float}
\usepackage{subcaption}

\usepackage[framemethod=default]{mdframed}

\newmdenv[backgroundcolor=gray!15,%
skipabove=5pt,%
skipbelow=5pt,%
leftmargin=2pt,%
rightmargin=2pt,%
innertopmargin=-6pt,%
innerbottommargin=5pt,%
innerleftmargin=5pt,%
innerrightmargin=5pt,%
splittopskip=0pt,%
splitbottomskip=0pt,%
linewidth=0pt,%
nobreak=true]%
{keyeqn}

\newmdenv[backgroundcolor=gray!15,%
skipabove=5pt,%
skipbelow=5pt,%
leftmargin=2pt,%
rightmargin=2pt,%
innertopmargin=-2pt,%
innerbottommargin=5pt,%
innerleftmargin=5pt,%
innerrightmargin=5pt,%
splittopskip=0pt,%
splitbottomskip=0pt,%
linewidth=0pt,%
nobreak=true]%
{keythrm}

\graphicspath{{Fig/}}

\definecolor{red2}{RGB}{214, 39, 40}
\definecolor{green2}{RGB}{0,170,0}
\definecolor{blue2}{RGB}{0,100,200}
\definecolor{magenta2}{RGB}{191,64,191}
\definecolor{purple2}{RGB}{112,48,160}
\definecolor{orange2}{RGB}{255,192,0}

\def\bgm{\begin{matrix}}
\def\edm{\end{matrix}}
\def \Mpl {M_{\text{Pl}}}

\numberwithin{equation}{section}

\allowdisplaybreaks[1]
\graphicspath{{Fig/}}

\begin{document}

\begin{titlepage}
\begin{flushright}
KYUSHU-HET-338
\end{flushright}

		\setcounter{page}{1} \baselineskip=15.5pt 
		\thispagestyle{empty}

\begin{minipage}{5cm}

\end{minipage}

	\begin{center}
		{\fontsize{18}{18}\centering {\bf{Heavy Field Effects on Inflationary Models\\[10pt] in Light of ACT Data}}}
	\end{center}
    
		\vskip 15pt
		\begin{center}
			\noindent
			{\fontsize{12}{18} \selectfont 
            Shuntaro Aoki$^{a,}$\footnote{shuntaro.aoki@riken.jp}, Hajime Otsuka$^{b,c,}$\footnote{otsuka.hajime@phys.kyushu-u.ac.jp}, and Ryota Yanagita$^{b,}$\footnote{ryota@phys.kyushu-u.ac.jp} 
   }

		\end{center}
		
		\begin{center}
			\vskip 8pt
			\textit{$^a$RIKEN Center for Interdisciplinary Theortical and Mathematical
Sciences (iTHEMS),
 Wako, Saitama 351-0198, Japan
}

	\textit{$^b$Department of Physics, Kyushu University, 744 Motooka, Nishi-ku, Fukuoka 819-0395, Japan}

	\textit{$^c$Quantum and Spacetime Research Institute (QuaSR), Kyushu University, 744 Motooka, Nishi-ku, Fukuoka 819-0395, Japan}
        \end{center}

\noindent\rule{\textwidth}{0.4pt}
 \begin{center}
     \noindent \textbf{Abstract}
 \end{center}
Recent results from the Atacama Cosmology Telescope (ACT), when combined with Planck and DESI datasets, indicate a scalar spectral index $n_s$ larger than that reported in the Planck 2018 baseline, thereby challenging conventional Starobinsky-type ($\alpha$-attractor) inflationary scenarios at the $2\sigma$ level. In addition, the positive running of the spectral index $\alpha_s$ implied by the data provides strong constraints on these models. In this paper, we explore the possibility that the presence of an additional heavy field during inflation, with a mass of order the Hubble scale and a sizable mixing coupling to the inflaton, can reconcile such inflationary models with the ACT results by increasing both $n_s$ and $\alpha_s$, particularly in the strong-mixing regime. Furthermore, we extend this framework to traditional inflation models such as chaotic inflation and natural inflation, which have already been excluded by Planck alone, and show that they can be revived in certain regions of parameter space. Inflationary observables, including the spectral index $n_s$, the tensor-to-scalar ratio $r$, and the running $\alpha_s$, are computed within the single-field EFT approach, which is applicable even in the presence of a heavy field with large mixing. We also discuss the non-Gaussianity signatures arising from the heavy field, noting that parts of the parameter space are already excluded or can be tested in future observations. Finally, we present concrete model realizations that allow for such a large mixing.    
	
		
	\end{titlepage}


\newpage
\setcounter{page}{2}
{
	\tableofcontents
}

\newpage



\section{Introduction}
The recent ACT data release~\cite{ACT:2025fju,ACT:2025tim}, in combination with DESI~\cite{DESI:2024mwx} and Planck~\cite{Planck:2018jri} data, imposes tighter constraints on the spectral index $n_s$ and the running of the spectral index~$\alpha_s$:
\begin{align}
n_s=0.9743\pm 0.0034\,(68\% \mathrm{CL}),\quad \alpha_s \equiv \mathrm{d} n_s / \mathrm{d} \ln k=0.0062 \pm 0.0052 \quad(68 \% \mathrm{CL}),\label{ACT}
\end{align}
which disfavors Starobinsky-type inflation models~\cite{Starobinsky:1980te, Bezrukov:2007ep, Kallosh:2013yoa} whose typical predictions are
\begin{align}
n_s =1-\frac{2}{N}\approx 0.965,\quad \alpha_s=-\frac{2}{N^2}\approx-0.0006
\end{align}
for $N \approx 60$ e-folds at the CMB pivot scale. There have been several attempts to explain the ACT results and to reconcile these models with observations~\cite{Kallosh:2025rni,Aoki:2025wld,Brahma:2025dio,Dioguardi:2025vci,Berera:2025vsu,Gialamas:2025kef,Salvio:2025izr,Dioguardi:2025mpp,Gao:2025onc,He:2025bli,Drees:2025ngb,Zharov:2025evb,Haque:2025uri,Liu:2025qca,Yin:2025rrs,Gialamas:2025ofz,Haque:2025uis,Yogesh:2025wak,Yi:2025dms,Addazi:2025qra,Maity:2025czp,Peng:2025bws,Mondal:2025kur,Frolovsky:2025iao,Haque:2025uga,Pallis:2025nrv,Chakraborty:2025oyj,Odintsov:2025wai,Heidarian:2025drk,Choudhury:2025vso,Okada:2025lpl,Gao:2025viy,Han:2025cwk,Pallis:2025gii,GonzalezQuaglia:2025qem,Zharov:2025zjg,Das:2025bws,Lynker:2025wyc,Zahoor:2025nuq,Ye:2025idn,German:2025ide,Ketov:2025cqg,Zhu:2025twm,Odintsov:2025bmp,Ellis:2025ieh,Yuennan:2025kde,Oikonomou:2025htz,Odintsov:2025jky,Oikonomou:2025xms,Cheong:2025vmz}, although some caution has been raised~\cite{Ferreira:2025lrd,Linde:2025pvj}.

In this paper, we point out another possibility for reconciling Starobinsky-type inflation with current observational results: the existence of an additional scalar field with a Hubble-scale mass (denoted as $m$). In general, this field can couple to the inflaton at the quadratic level of perturbations in the de Sitter background, and it can also have linear mixing (denoted as $\rho$) with inflaton fluctuations. Such an inflationary scenario is referred to as quasi-single field inflation, originally proposed in Ref.~\cite{Chen:2009zp} for the case with small mixing coupling. In particular, scalar fields with Hubble-scale masses are ubiquitous in inflationary models based on supergravity~\cite{Stewart:1994ts}. 

The effects of such heavy fields on inflationary observables (the spectral index $n_s$ and the tensor-to-scalar ratio $r$) were studied in detail in Ref.~\cite{Tong:2017iat} using a refined version of single-field EFT, which is applicable not only to the case of a Hubble mass but also to a large mixing coupling. In this work, we also apply the formalism to the running of the spectral index $\alpha_s$, which serves as another important indicator, as mentioned above. We find that, for certain regions of parameter space for $m$ and $\rho$, the Starobinsky-type models can be reconciled with ACT data. We further apply the same method to chaotic inflation and natural inflation.

Interestingly, in these parameter regions, the non-Gaussianity induced by the heavy field is enhanced and can sometimes have a characteristic oscillatory feature, providing a unique observational test of this scenario and allowing it to be distinguished from others. 

While the mass and mixing couplings are treated as free parameters in the single-field EFT approach, it remains an open question whether they can be realized in concrete models. To address this, we consider some simple realizations of quasi-single field inflation with (large) linear mixing, starting from a two-field system.

The paper is organized as follows. In Sec.~\ref{sec:EFT}, we review the single-field EFT method used to derive inflationary observables, and slightly update it including the running of the spectral index. We then apply this framework to various inflation models: Starobinsky-type, chaotic, and natural inflation. In Sec.~\ref{sec:NG}, we discuss non-Gaussianity in our setup. In Sec.~\ref{sec:RTI}, we consider concrete model realizations in a two-field system. Finally, Sec.~\ref{sec:summary} is devoted to the summary.

\section{Inflationary Observables with an Additional Massive Field}\label{sec:EFT}
Here we compute the inflationary observables (in particular, the spectral index $n_s$, the tensor-to-scalar ratio $r$, and the running of the spectral index $\alpha_s$) for a quasi-single-field inflation setup, i.e., one light field corresponding to the inflaton and a heavy field with mass around the Hubble scale. We employ the single-field effective field theory (EFT) obtained by integrating out the heavy mode, which induces a non-trivial sound speed. We then apply the resulting formulae to several types of inflaton scalar potentials. 

Most of the corresponding results can be found in Ref.~\cite{Tong:2017iat}, where the authors compare them with the Planck 2018 data~\cite{Planck:2018jri, Planck:2019kim}. Here we update the analysis by confronting it with the more recent ACT results~\cite{ACT:2025fju,ACT:2025tim}, including a new calculation of $\alpha_s$.

\subsection{Inflationary observables based on single-field EFT}
We start from a two-field system with $\phi^a\ (a=1,2)$, consisting of the inflaton and an additional scalar field, with the generic action
\begin{align}
S=\int d^4 x \sqrt{-g}\left[\frac{\Mpl^2}{2}R-\frac{1}{2} G_{ab} (\phi)\partial_\mu \phi^a \partial^\mu \phi^b-V\left(\phi\right)\right],   \label{L_2_2}
\end{align}
where $G_{ab} (\phi)$ and $V\left(\phi\right)$ denote the field-space metric and the scalar potential, respectively. The cosmological perturbations in this setup have been discussed in several works, e.g., Refs.~\cite{Sasaki:1995aw, GrootNibbelink:2000vx, GrootNibbelink:2001qt, Langlois:2008mn, Langlois:2008qf, Peterson:2010np, Gong:2011uw, Elliston:2012ab, Kaiser:2012ak, Garcia-Saenz:2019njm}, and see also Refs.~\cite{Wang:2013zva, Gong:2016qmq} for reviews. 
Here we only present the main results, leaving further details to Appendix~\ref{app:two}.

The fluctuations of the two scalar fields can be decomposed into the massless curvature perturbation $\zeta$ and the massive isocurvature perturbation $\sigma$ in the FLRW background 
\[
ds^2 = -dt^2 + a(t)^2 dx^2 ,
\]
with scale factor $a(t)$ and homogeneous inflaton background $\phi^a_0(t)$. At the quadratic level, the action takes the form
\begin{align}
S[\zeta, \sigma]=\int dt d^3 x \ a^3\left[\frac{\dot{\phi}_0^2}{2 H^2}\left(\dot{\zeta}^2-\frac{\left(\partial_i \zeta\right)^2}{a^2}\right)+\frac{1}{2}\left(\dot{\sigma}^2-\frac{\left(\partial_i \sigma\right)^2}{a^2}-m^2 \sigma^2\right)- 2\eta_\perp \dot{\phi}_0 \dot{\zeta} \sigma\right],    \label{S_quad_a}
\end{align}
where the dot denotes a time derivative $d/dt$, $\partial_i$ represents the spatial derivative, and $H \equiv \dot{a}/a$ is the Hubble parameter. We define the total velocity of the background fields $\dot{\phi}_0^2 \equiv G_{ab} \dot{\phi}_0^a \dot{\phi}_0^b$. The parameter $m$ denotes the mass of the isocurvature perturbation $\sigma$ (often referred to as the isocurvature mass), and $\eta_\perp$ characterizes the linear mixing between $\zeta$ and $\sigma$ (called the turn rate). Both quantities are determined by the background dynamics as
\begin{align}
\eta_{\perp} = \frac{V_N}{\dot{\phi}_0 H}, \qquad 
m^2 = V_{NN} - \eta_\perp^2 H^2 + \epsilon H^2 M_{\mathrm{Pl}}^2 \mathcal{R},
\end{align}
where $V_N \equiv N^a V_a$ and $V_{NN} \equiv N^a N^b V_{ab}$, with $V_a = \partial_a V \equiv \partial V / \partial \phi^a$ and $V_{ab} \equiv \nabla_a V_b = \partial_a V_b - \Gamma^c_{ab} V_c$, are the derivatives of the potential projected along the normal vector $N_a$ (and tangent vector $T^a$) to the background trajectory, defined by
\begin{align}
T^a = \frac{\dot{\phi}_0^a}{\dot{\phi}_0}, \qquad 
N_a = (\det G)^{1/2} \epsilon_{ab} T^b,
\end{align}
with $\epsilon_{12} = -\epsilon_{21} = 1$ and $\epsilon_{11} = \epsilon_{22} = 0$. 
Field-space indices are raised and lowered using the metric $G_{ab}$ and its inverse $G^{ab}$. 
Here $\epsilon \equiv -\dot{H}/H^2$ is the first slow-roll parameter, and $\mathcal{R}$ is the Ricci scalar constructed from the field-space metric $G_{ab}$. As noted in Ref.~\cite{Achucarro:2010da}, the turn rate $\eta_\perp$ can also be expressed as
\begin{align}
|\eta_\perp| = \sqrt{2\epsilon}\, M_{\mathrm{Pl}}\, \kappa,\label{eta_perp_a}
\end{align}
where $\kappa$ represents the curvature of the background trajectory, which is independent of the field velocity. 
In the following, we introduce the notation
\begin{align}
\rho \equiv 2 \eta_\perp H.   
\end{align}

Now following Ref.~\cite{Tong:2017iat}, we consider the single-field EFT for the curvature perturbation $\zeta$ by integrating out the isocurvature~$\sigma$, which is valid as long as $m^2+\rho^2>4 H^2$ is satisfied. We focus on an inflationary (approximately de Sitter) background with $a = e^{Ht}$, where $H \simeq \mathrm{const.}$, or equivalently $a = -1/(H\tau)$ in terms of the conformal time $\tau$ defined by $dt = a\, d\tau$. The equation of motion for the canonically normalized isocurvature perturbation $\widetilde{\sigma} \equiv a\sigma$ is then given in momentum space with $k$ by
\begin{align}
\square \widetilde{\sigma} = \frac{a^2 \dot{\phi}_0 \rho}{H} \zeta^{\prime},
\quad \mathrm{where} \quad
\square \equiv -\partial_\tau^2 - k^2 - m^2 a^2 + 2 a^2 H^2,
\end{align}
and the prime denotes differentiation with respect to $\tau$. This equation can be solved as
\begin{align}
\widetilde{\sigma}=\frac{-1}{k^2+a^2 m^2 -2 a^2 H^2}\frac{a^2\dot{\phi}_0\rho}{H}\zeta^{\prime}+\cdots,\label{sigma_int}
\end{align}
where the ellipses denotes the terms including higher-order time derivative on $\zeta$. We omit them,\footnote{This constrains the regime of validity of the EFT. As shown in Ref.~\cite{Garcia-Saenz:2019njm}, for the EFT to be consistent, the adiabaticity conditions must be satisfied, meaning that the background quantities $H$, $\epsilon$, $\rho$, and $m$ evolve on time scales much longer than $m^{-1}$, which we assume throughout this paper.} and by inserting only the first term back into the original action and consistently neglecting the time derivative, we obtain the single-field EFT for $\zeta$,
\begin{align}
S_{\mathrm{eff}}[\zeta]= \int d \tau\,\frac{d^3 k}{(2 \pi)^3}\, a^2\Mpl^2 \epsilon\left[\left(1+\frac{\rho^2}{k^2 H^2 \tau^2+m^2-2 H^2}\right) \zeta^{\prime 2}-k^2 \zeta^2\right],
\end{align}
from which one can read off the effective sound speed $c_s$ as
\begin{align}
c_s^{-2}(k \tau)=1+\frac{\rho^2}{k^2 H^2 \tau^2+m^2-2 H^2}.  \label{c_swithk}  
\end{align}
The large-$m$ limit and the large-$\rho$ limit of $c_s$ have been discussed in Refs.~\cite{Tolley:2009fg, Achucarro:2010jv, Achucarro:2012sm}, and~\cite{Cremonini:2010ua, Baumann:2011su, Gwyn:2012mw, An:2017hlx}, respectively. In Ref.~\cite{Tong:2017iat}, the authors provided a unified treatment that incorporates both limits by employing a horizon-crossing approximation and matching to the result in the large-$\rho$ limit.\footnote{More specifically, Ref.~\cite{Tong:2017iat} sets $-k\tau = B\,c_s^{-1}$ with a constant $B$ on the right-hand side of Eq.~\eqref{c_swithk}, which yields an equation that can be solved for $c_s$:
\begin{align}
c_s^{-2}=1+\frac{\rho^2}{H^2 B^2 c_s^{-2}+m^2-2 H^2}.    
\end{align}
The explicit solution is given by
\begin{align}
c_s^{-1}=\sqrt{\frac{2\left(\rho^2+m^2-2 H^2\right)}{\sqrt{4 B^2 H^2 \rho^2+\left(B^2 H^2+m^2-2 H^2\right)^2}-B^2 H^2+m^2-2 H^2}}.   
\end{align}
Then, by matching this to the known result 
$c_s^{-1}=\mathcal{C}\sqrt{\rho/H}$ for $\rho H\gg m^2$ obtained in Refs.~\cite{Gwyn:2012mw,An:2017hlx}, 
the parameter $B$ is fixed as $B=\mathcal{C}^{-2}$ (where $\mathcal{C}$ is defined in the main text), 
which reproduces the result shown in Eq.~\eqref{c_s}. 

This procedure is justified by noting that the curvature perturbation $\zeta$ freezes out at the time scale $-k\tau \sim c_s^{-1}$. 
Numerical comparison between the solutions with Eq.~\eqref{c_swithk} and Eq.~\eqref{c_s} shows good agreement, 
with only small deviations as long as $m^2+\rho^2\gtrsim 2H^2$~\cite{Tong:2017iat}.
} The result is
\begin{align}
c_s^{-1}=\sqrt{\frac{2\left(\tilde{m}^2-2 +\tilde{\rho}^2\right)}{\tilde{m}^2-2 -\frac{1}{\mathcal{C}^4}+\sqrt{\left(\tilde{m}^2-2 +\frac{1}{\mathcal{C}^4}\right)^2+\frac{4  \tilde{\rho}^2}{\mathcal{C}^4}}}}, \label{c_s}   
\end{align}
where $\tilde{\rho}\equiv \rho/H$ and $\tilde{m}\equiv m/H$ are the dimensionless mixing parameter and mass, and $\mathcal{C}=16\pi/(\Gamma[-1/4])\sim 2.09$ is a number. It interpolates the large mass and large mixing expressions as
\begin{align}
c_s^{-1}\simeq \left\{\begin{array}{l}
\sqrt{1+\frac{\tilde{\rho}^2}{\tilde{m}^2-2 }}\quad {\rm{for}} \quad m^2 \gg \rho H  \\
\mathcal{C}\sqrt{\tilde{\rho}}  \quad {\rm{for}} \quad m^2 \ll \rho H 
\end{array}\right.
\end{align}
by construction.

Based on $c_s$ in Eq.~\eqref{c_s}, the power spectrum is modified from the standard single field theory as
\begin{align}
P_\zeta=\frac{H^2}{8 \pi^2 c_s \epsilon \Mpl^2}.   \label{P_zeta} 
\end{align}
Accordingly, the spectral index $n_s$ is corrected and evaluated as~\cite{Tong:2017iat}
\begin{align}
n_s-1 \equiv \frac{d \ln P_\zeta}{d \ln k}=-2 \epsilon-\eta+\frac{\partial \ln c_s^{-1}}{\partial \ln \tilde{\rho}} \frac{\eta}{2} +\frac{\partial \ln c_s^{-1}}{\partial \ln \tilde{m}} \epsilon,\label{n_s}
\end{align}
where $\epsilon$ and $\eta$ are the slow-roll (SR) parameters defined by
\begin{align}
\epsilon\equiv -\frac{\dot{H}}{H^2}, \quad \eta\equiv\frac{\dot{\epsilon}}{\epsilon H}. \label{def_SR}
\end{align}
In the derivation, we used the fact $\tilde{\rho}\propto \sqrt{\epsilon}$, see Eq.~\eqref{eta_perp_a}.
Since the tensor power spectrum $P_\gamma=2H^2/(\pi^2\Mpl^2)$ is not affected by the  existence of isocurvature mode~$\sigma$, the tensor-to-scalar ratio $r$ is given by
\begin{align}
r\equiv \frac{P_\gamma}{P_\zeta}=16 c_s \epsilon.
\end{align}
In this work, we are also interested in the running of the spectral index $\alpha_s$ which can be computed by
\begin{align}
\nonumber \alpha_s \equiv&\ \frac{d \ln n_s}{d \ln k}\\
\nonumber=&\ -2\epsilon \eta -\eta \eta_2+\frac{\partial^2 \ln c_s^{-1}}{\partial^2 \ln \tilde{\rho}}  \frac{\eta}{2}+\frac{\partial^2 \ln c_s^{-1}}{\partial \ln \tilde{\rho}\ \partial \ln \tilde{m}} \left(\epsilon+\frac{\eta}{2}\right)+\frac{\partial^2 \ln c_s^{-1}}{\partial^2 \ln \tilde{m}}\epsilon\\
&+ \frac{\partial \ln c_s^{-1}}{\partial \ln \tilde{\rho}} \frac{\eta\eta_2}{2}+\frac{\partial \ln c_s^{-1}}{\partial \ln \tilde{m}} \epsilon \eta,\label{a_s}
\end{align}
where we have introduced the third SR parameter by
\begin{align}
\eta_2\equiv\frac{\dot{\eta}}{H\eta}. \label{def_SR2}    
\end{align}

The expressions~\eqref{n_s} and~\eqref{a_s} are complicated in general (see Appendix~\ref{app:full} for their explicit expressions), but their leading behavior in the limits of large mass or strong mixing can be extracted as follows:
\begin{itemize}
\item {\bf Heavy mass limit: $m^2 \gg \rho H$}
\begin{align}
&n_s = 1 - 2 \epsilon - \eta +\frac{\tilde{\rho}^2}{2\tilde{m}^2}\left(\eta-2\epsilon \right) + \cdots,\\
&\alpha_s = -2 \epsilon \eta - \eta \eta_2+\frac{\tilde{\rho}^2}{2\tilde{m}^2}\left(\eta_2-2\epsilon\right)\eta+ \cdots,
\end{align}
\item {\bf Strong mixing limit: $m^2 \ll \rho H$}
\begin{align}
&n_s = 1 - 2 \epsilon - \tfrac{3}{4}\eta + \cdots,\label{n_s_rho}\\
&\alpha_s = -2 \epsilon \eta - \tfrac{3}{4}\eta \eta_2 + \cdots.\label{a_s_rho}
\end{align}
\end{itemize}
As noted in Ref.~\cite{Tong:2017iat} for the case of $n_s$, the running $\alpha_s$ likewise approaches universal values independent of $m$ in the strong mixing regime, Eqs.~\eqref{n_s_rho} and~\eqref{a_s_rho}, since $c_s^{-1} \propto \tilde{\rho}^{1/2}$. In particular, strong mixing tends to increase both $n_s$ and $\alpha_s$, which play an important role in allowing several single-field inflation models to be reconciled with the ACT data.

We now have the formulae for $n_s$, $r$, and $\alpha_s$ at hand, which depend on the SR parameters $\epsilon, \eta, \eta_2$ as well as on the mass $m$ and the mixing parameter $\rho$ through the sound speed $c_s(m,\rho)$. Since the SR parameters are determined by the form of the inflaton potential, we turn to specific models in the next subsection.

\subsection{Impacts on several inflation models}
\label{sec:inflation models}
Before turning to specific models, let us first introduce the so-called potential slow-roll (SR) parameters, defined in terms of derivatives of the inflaton potential as
\begin{align}
\epsilon_V \equiv \frac{1}{2}\left(\frac{V^{\prime}}{V}\right)^2, \quad 
\eta_V \equiv  \frac{V^{\prime \prime}}{V}, \quad 
\xi_V \equiv \frac{V^{\prime} V^{\prime \prime \prime}}{V^2},
\end{align}
where the prime denotes a derivative with respect to inflaton $\phi$. We set the Planck scale to unity, $\Mpl = 1$, here and throughout unless stated otherwise. These potential SR parameters are related to the original SR parameters in Eqs.~\eqref{def_SR} and~\eqref{def_SR2} by
\begin{align}
\epsilon \simeq \epsilon_V,\quad 
\eta \simeq -2\eta_V + 4\epsilon_V,\quad 
\eta_2 \simeq \frac{8\epsilon_V^2 - 6\epsilon_V \eta_V - 2\eta_V^2 + \xi_V}{2\epsilon_V - \eta_V},
\end{align}
at leading order.

Therefore, the formulae~\eqref{n_s} and~\eqref{a_s} can be rewritten in terms of the potential SR parameters as
\begin{align}
n_s \simeq &\ 1 - 6 \epsilon_V + 2 \eta_V 
+ \frac{\partial \ln c_s^{-1}}{\partial \ln \tilde{\rho}} (2\epsilon_V - \eta_V) 
+ \frac{\partial \ln c_s^{-1}}{\partial \ln \tilde{m}} \epsilon_V, \label{n_s_2}\\
\nonumber \alpha_s \simeq &\ -24 \epsilon_V^2 + 16 \epsilon_V \eta_V - 2 \xi_V 
+ \frac{\partial^2 \ln c_s^{-1}}{\partial^2 \ln \tilde{\rho}} (2\epsilon_V - \eta_V) 
+ \frac{\partial^2 \ln c_s^{-1}}{\partial \ln \tilde{\rho}\ \partial \ln \tilde{m}} (3\epsilon_V - \eta_V) \\
&+ \frac{\partial^2 \ln c_s^{-1}}{\partial^2 \ln \tilde{m}} \epsilon_V
+ \frac{\partial \ln c_s^{-1}}{\partial \ln \tilde{\rho}} (8 \epsilon_V^2 - 6 \epsilon_V \eta_V - 2\eta_V^2 + \xi_V) 
+ \frac{\partial \ln c_s^{-1}}{\partial \ln \tilde{m}} (4 \epsilon_V^2 - 2\epsilon_V \eta_V). \label{a_s_2}
\end{align}
Note the above equations reproduce the standard results when $c_s = 1$~\cite{Zarei:2014bta}.

We now apply the formulae developed in the previous subsection to some representative inflation models:
\begin{align}
V(\phi) \propto 
\begin{cases}
1 - e^{-\sqrt{\frac{2}{3\alpha}}\,\phi} : & \text{Starobinsky-type inflation} \\
\phi^n : & \text{Chaotic inflation} \\
1 - \cos\frac{\phi}{f} : & \text{Natural inflation}
\end{cases}
,
\end{align}
where $\alpha$, $n$, and $f$ are real parameters characterizing these models. The Starobinsky model~\cite{Starobinsky:1980te} originating from higher-curvature gravity corresponds to $\alpha = 1$, while more general choices of $\alpha$ correspond to the $\alpha$-attractor models~\cite{Kallosh:2013yoa}. Higgs inflation~\cite{Bezrukov:2007ep} takes almost the same form in the large field regime. We also discuss the effects of a massive field on chaotic inflation~\cite{Linde:1983gd} and natural inflation~\cite{Freese:1990rb}.

In the standard scenario without an additional heavy field (or in the case where it is completely decoupled, i.e., $m/H \gg 1$ or $c_s = 1$), it is straightforward to calculate the potential SR parameters and the inflationary observables as functions of the number of e-folding $N$ (see Ref.~\cite{Baumann:2009ds} for review), which are summarized in Table~\ref{tab:model}. These values should be understood as being at leading order in $1/N$ for the Starobinsky-type and chaotic inflation models. We take $f \gtrsim 1$ for the case of natural inflation,\footnote{This is the well-known trans-Planckian problem in natural inflation~\cite{Banks:2003sx}: Successful natural inflation requires a trans-Planckian decay constant, which raises questions about the validity of the EFT. The Kim-Nilles-Peloso (KNP) alignment mechanism~\cite{Kim:2004rp} provides a possible solution based on the presence of two axions. See, e.g., Refs.~\cite{Rudelius:2015xta,Montero:2015ofa,Brown:2015iha,Junghans:2015hba,Heidenreich:2015nta}, for the constraints from the weak gravity conjecture~\cite{Arkani-Hamed:2006emk} for the aligned natural inflation. It is not clear whether the heavy field direction in the alignment mechanism can be identified with our additional massive field, but we leave this as an interesting question for future work, particularly in relation to the UV embedding of our scenario.} but do not impose this condition when plotting the figures discussed below.

\begin{table}[h]
\centering
\caption{SR parameters and inflationary observables with $c_s = 1$.}
\label{tab:model}
\renewcommand{\arraystretch}{1.5}
\begin{tabular}{ccccccc}
\toprule
\textbf{Model} & \( \epsilon_V \) & \( \eta_V \) & \( \xi_V \) & \( r \) & \( n_s-1 \) & \( \alpha_s \) \\
\midrule
   \(1-e^{\sqrt{\frac{2}{3\alpha}}\phi}\) & \( \frac{3\alpha}{4N^2}  \)& \(-\frac{1}{N}\) &\(\frac{1}{N^2}\) &\( \frac{12\alpha}{N^2}  \) & \(-\frac{2}{N}\)&\(-\frac{2}{N^2}\) \\
   
    \(\phi^n\) & \( \frac{n}{4N}\)& \( \frac{n-1}{2N}\)& \( \frac{(n-1)(n-2)}{4N^2}\)& \( \frac{4n}{N}\) &\(-\frac{n+2}{2N}\) & \(-\frac{n+2}{2N^2}\)\\
   
   \(1-\cos \frac{\phi}{f}\) & \(\frac{e^{-N/f^2}}{2f^2} \)&\(-\frac{1}{2f^2}\)&\(-\frac{e^{-N/f^2}}{f^4}\)&\(\frac{8e^{-N/f^2}}{f^2}\)&\(-\frac{1}{f^2}\)& \(-\frac{2e^{-N/f^2}}{f^4}\) \\
\bottomrule
\end{tabular}
\end{table}

For the typical values of parameters and $N=60$, we obtain
\begin{align}
 (n_s,r,\alpha_s)=\begin{cases}
(0.965,0.0033,-0.0006),  & \alpha=1, \\
(0.967,0.13,-0.00055), & n=2, \\
(0.952,0.031,-0.003), & f=5,
\end{cases}
\end{align}
which can be compared with the ACT results~\eqref{ACT}. In particular, the positive value of $\alpha_s$ implied by the ACT data appears difficult to achieve, as pointed out in Ref.~\cite{Frolovsky:2025iao}.

\begin{figure}[H]
 \begin{minipage}{0.5\hsize}
  \begin{center}
   \includegraphics[width=80mm]{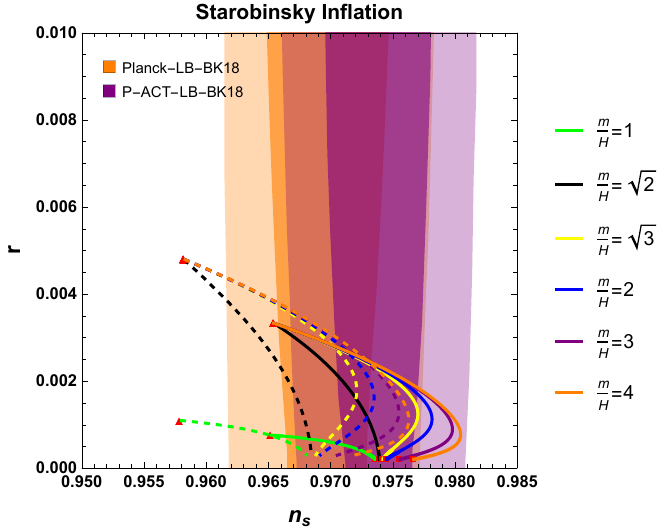}
  \end{center}
 \end{minipage}
 \begin{minipage}{0.5\hsize}
  \begin{center}
   \includegraphics[width=80mm]{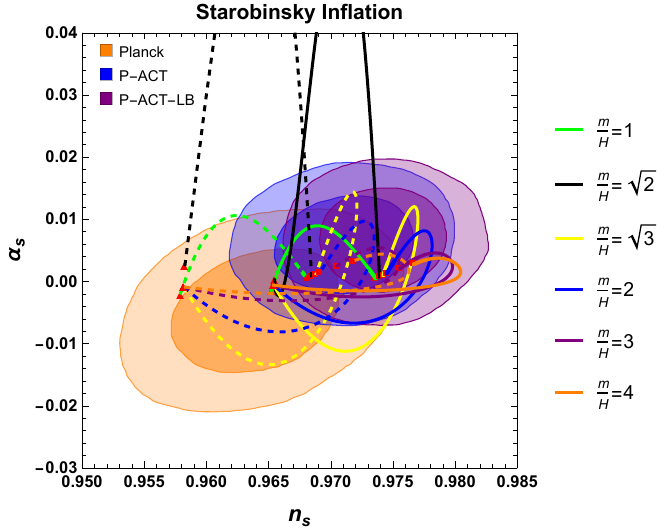}
  \end{center}
 \end{minipage}
\caption{Inflationary predictions in Starobinsky inflation with an additional field of mass $m$ and mixing coupling $\rho$. 
{\it Left} : $(n_s,r)$ plane obtained by varying $\rho$ while fixing $m$ as indicated in the panel. 
Solid (dashed) curves correspond to $N=60$ ($N=50$). Along each curve, $\rho/H$ is varied from $\rho_{\rm min}$ to $\rho_{\mathrm{max}}$, 
with markers at the edge $\blacktriangle$ and $\blacksquare$ denoting $\rho_{\rm min}$ and $\rho_{\mathrm{max}}$, respectively. 
Here $\rho_{\rm min}$ is chosen as the minimal value of $\rho$ consistent with the validity of the EFT description, $m^2+\rho^2>4H^2$, whereas $\rho_{\mathrm{max}}$ is fixed as $\rho_{\mathrm{max}}/H= 61,64,66,68,77,89$ for $m/H=1,\sqrt{2},\sqrt{3},2,3,4$ to be consistent with the Planck constraints on the equilateral non-Gaussianity, $f_{\mathrm{NL}}^{\mathrm{eq}}=-26\pm 47$~\cite{Planck:2019kim}, as discussed later in Sec.~\ref{sec:NG}. 
The constraints on $(n_s,r)$ are derived from the combined Planck, ACT, and DESI data set (P-ACT-LB-BK18), 
with dark and light purple regions corresponding to the $1\sigma$ and $2\sigma$ confidence levels, respectively. 
The constraints from Planck-LB-BK18 are shown in dark and light orange.  
{\it Right} : $(n_s,\alpha_s)$ plane for the same parameter variation as in the left panel. 
The observational constraints are taken from Ref.~\cite{ACT:2025tim}, where Planck-only (orange), P-ACT (navy), 
and P-ACT-LB (purple) results are shown. }
  \label{fig1}
\end{figure}

\begin{figure}[H]
   \begin{minipage}{0.5\hsize}
  \begin{center}
   \includegraphics[width=80mm]{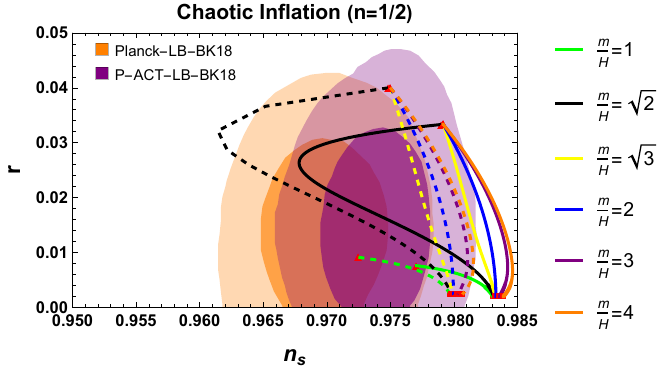}
  \end{center}
 \end{minipage}
 \begin{minipage}{0.5\hsize}
  \begin{center}
   \includegraphics[width=80mm]{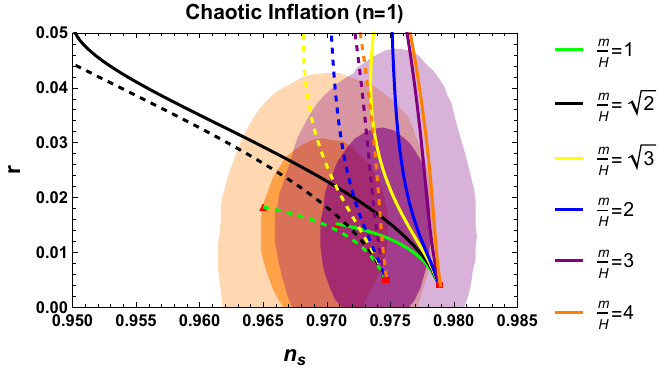}
  \end{center}
  \end{minipage}
     \begin{minipage}{0.5\hsize}
  \begin{center}
   \includegraphics[width=80mm]{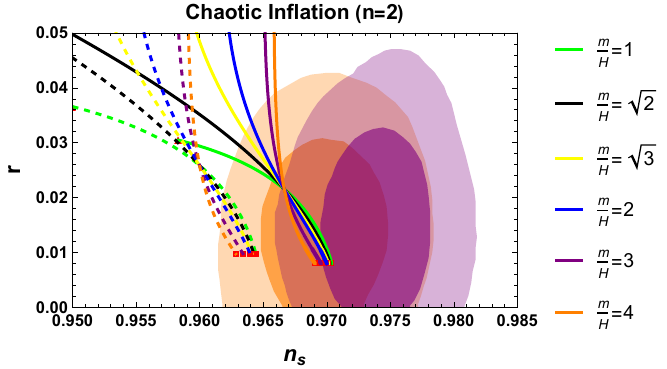}
  \end{center}
 \end{minipage}
    \begin{minipage}{0.5\hsize}
  \begin{center}
   \includegraphics[width=80mm]{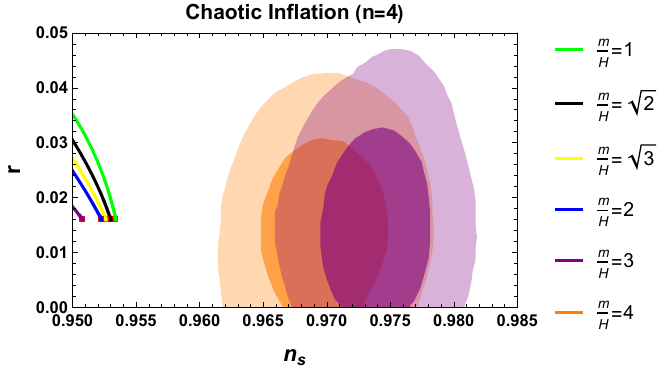}
  \end{center}
 \end{minipage}
   \caption{The same format as the left panel of Fig.~\ref{fig1}, but for chaotic inflation. 
From the upper left to the lower right, we show the $(n_s,r)$ plane for $n=\tfrac{1}{2},\,1,\,2,\,4$. }
    \label{fig2}
\end{figure}

\begin{figure}[H]
   \begin{minipage}{0.5\hsize}
  \begin{center}
   \includegraphics[width=75mm]{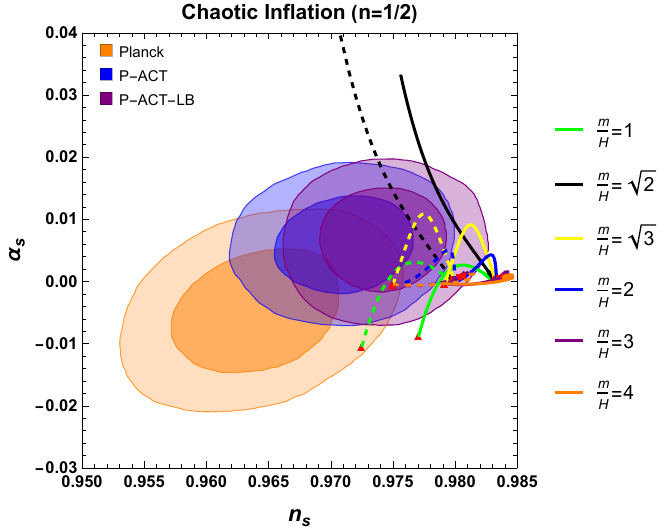}
  \end{center}
 \end{minipage}
 \begin{minipage}{0.5\hsize}
  \begin{center}
   \includegraphics[width=75mm]{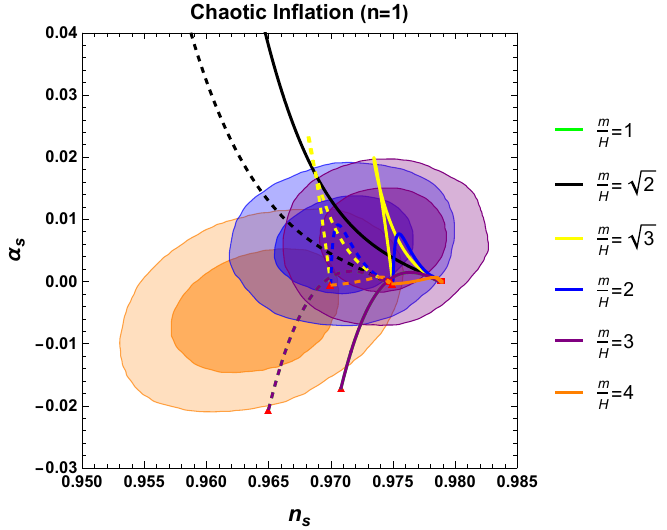}
  \end{center}
  \end{minipage}
     \begin{minipage}{0.5\hsize}
  \begin{center}
   \includegraphics[width=75mm]{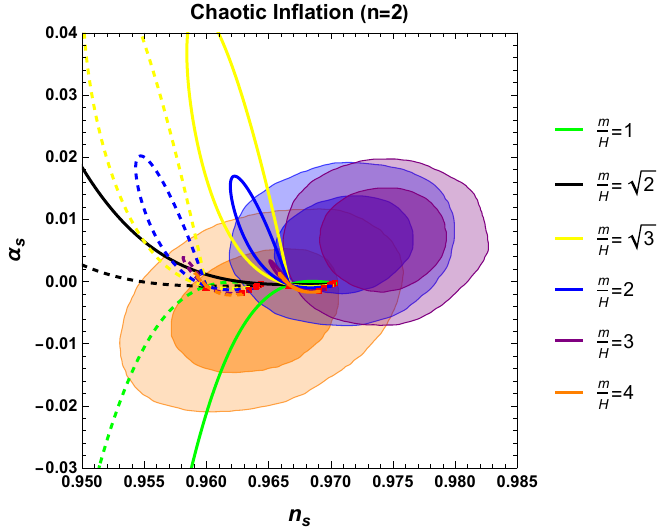}
  \end{center}
 \end{minipage}
    \begin{minipage}{0.5\hsize}
  \begin{center}
   \includegraphics[width=75mm]{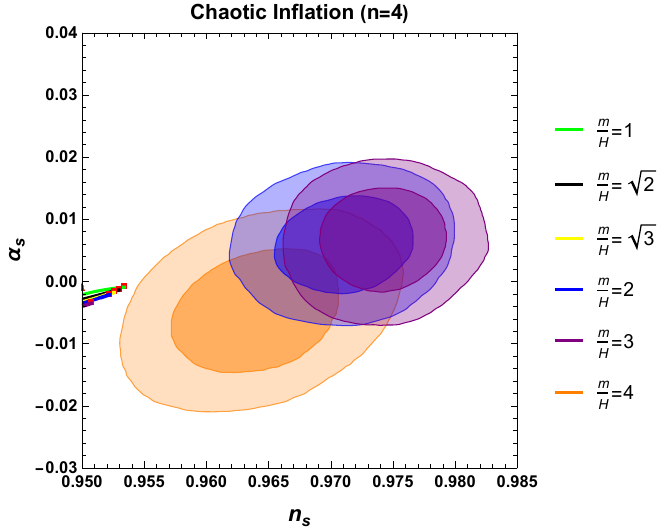}
  \end{center}
 \end{minipage}
   \caption{The same format as the right panel of Fig.~\ref{fig1}, but for chaotic inflation. 
From the upper left to the lower right, we show the $(\alpha_s,n_s)$ plane for $n=\tfrac{1}{2},\,1,\,2,\,4$. }
    \label{fig3}
\end{figure}

\begin{figure}[H]
 \begin{minipage}{0.5\hsize}
  \begin{center}
   \includegraphics[width=73mm]{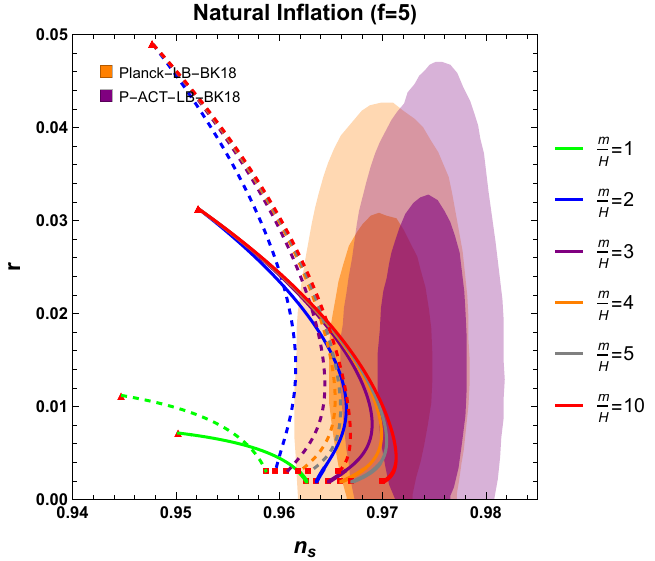}
  \end{center}
 \end{minipage}
 \begin{minipage}{0.5\hsize}
  \begin{center}
   \includegraphics[width=73mm]{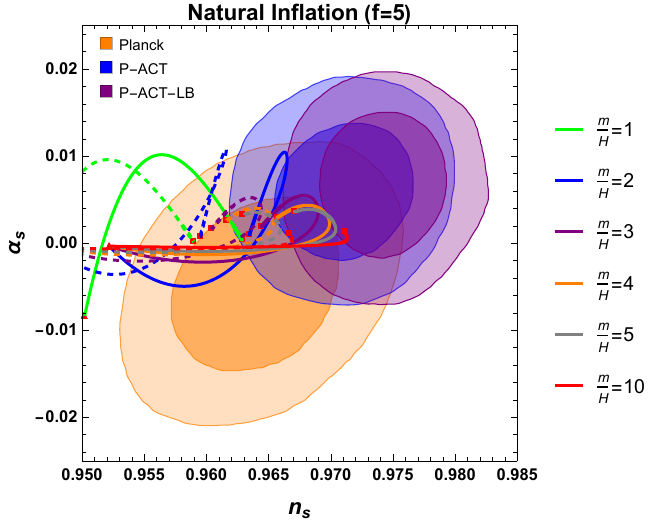}
  \end{center}
 \end{minipage}

\caption{Same format as  Fig.~\ref{fig1}, but for natural inflation. 
For $m/H=5,10$, $\rho_{\mathrm{max}}$ is taken as $\rho_{\mathrm{max}}/H= 102,177$ to be consistent with bound on non-Gaussianity.}
  \label{fig4}
\end{figure}

\begin{figure}[H]
\begin{minipage}{0.45\hsize}
  \begin{center}
   \includegraphics[width=72mm]{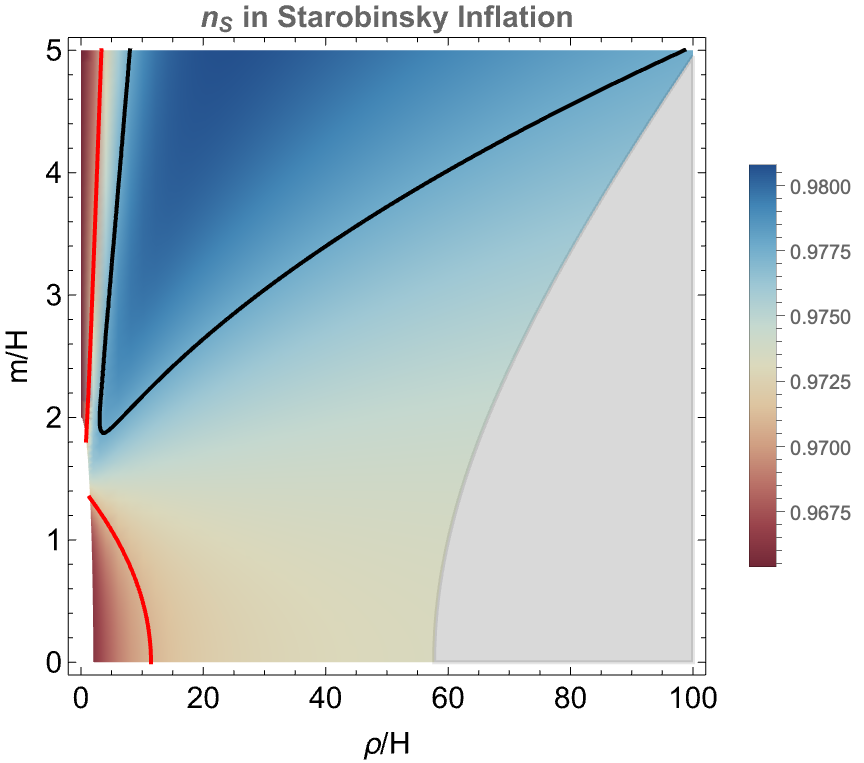}
  \end{center}
 \end{minipage}
\begin{minipage}{0.45\hsize}
  \begin{center}
   \includegraphics[width=72mm]{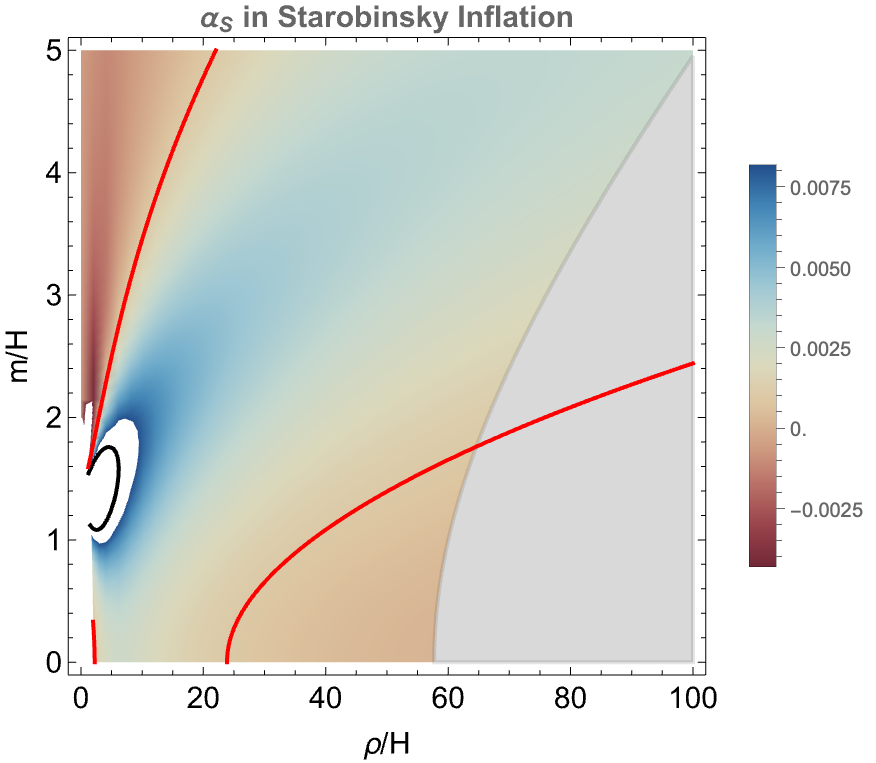}
  \end{center}
 \end{minipage}
 \begin{minipage}{0.45\hsize}
  \begin{center}
   \includegraphics[width=72mm]{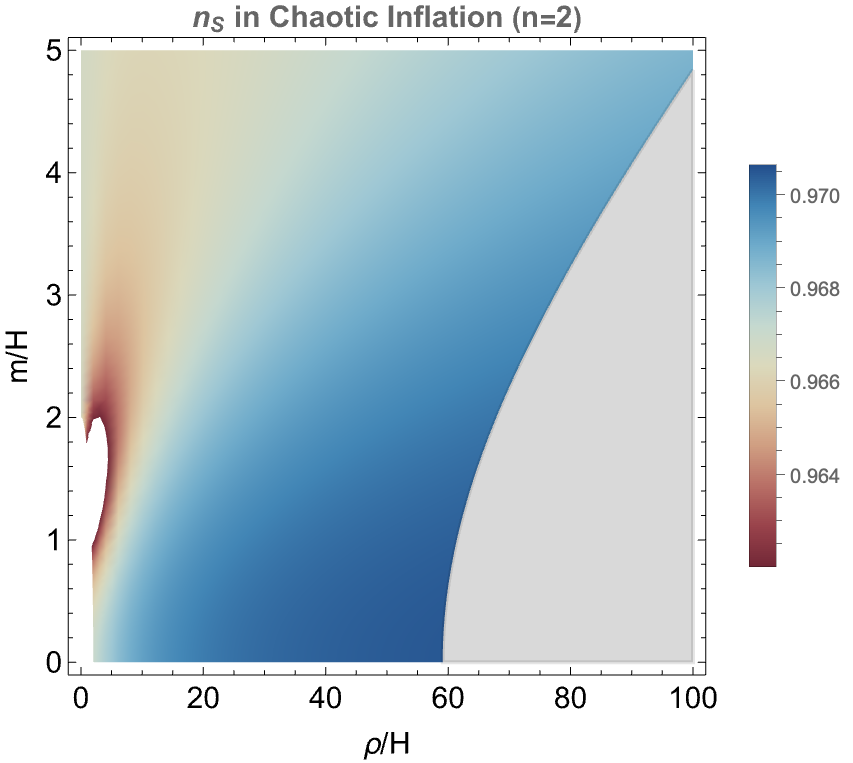}
  \end{center}
 \end{minipage}
 \begin{minipage}{0.45\hsize}
  \begin{center}
   \includegraphics[width=72mm]{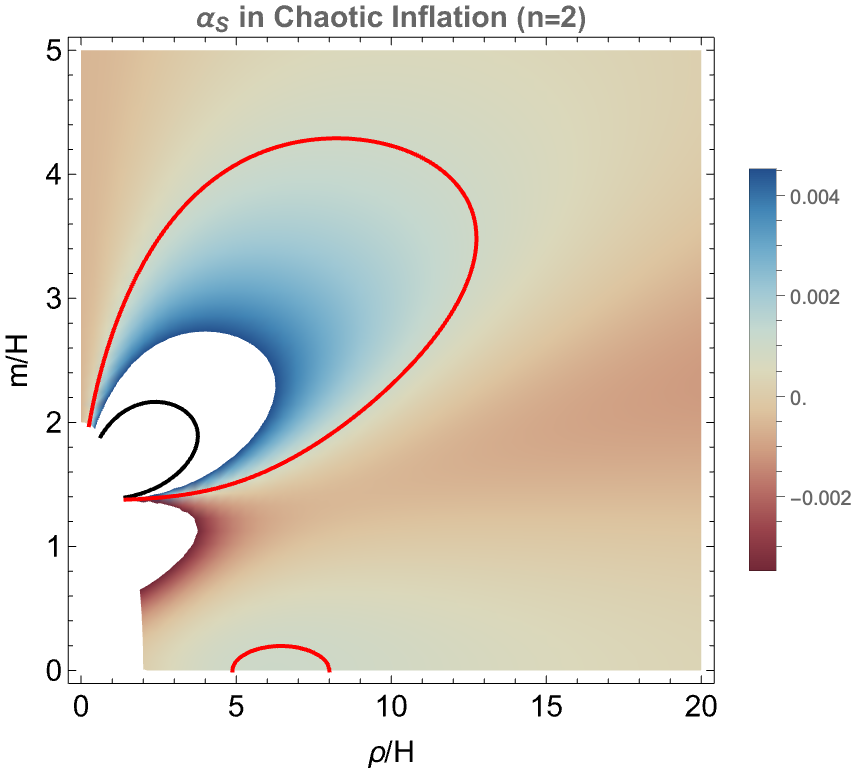}
  \end{center}
 \end{minipage}
 \begin{minipage}{0.45\hsize}
  \begin{center}
   \includegraphics[width=72mm]{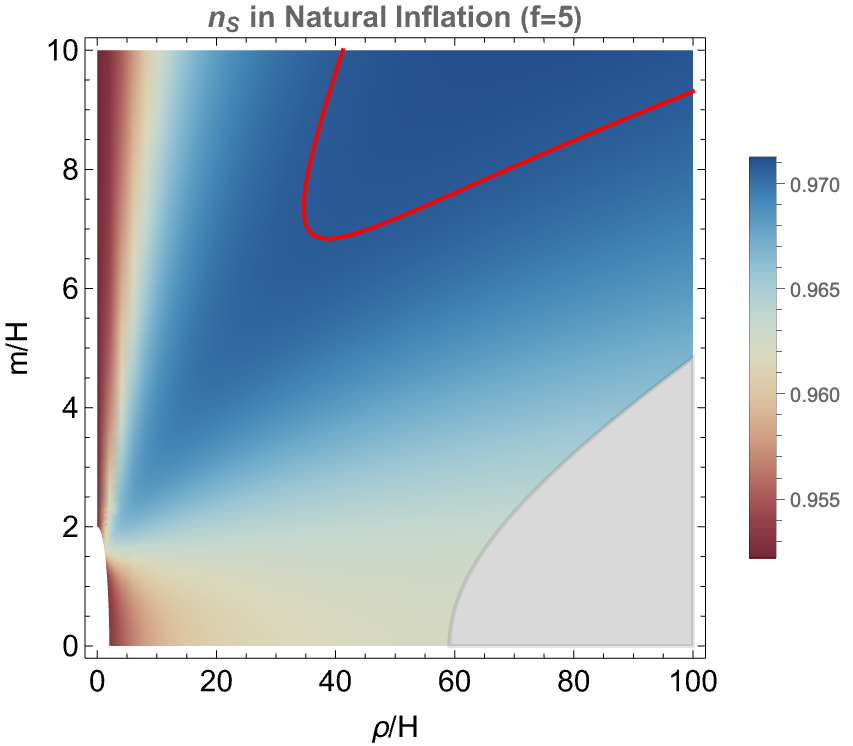}
  \end{center}
 \end{minipage}
 \begin{minipage}{0.65\hsize}
  \begin{center}
   \includegraphics[width=72mm]{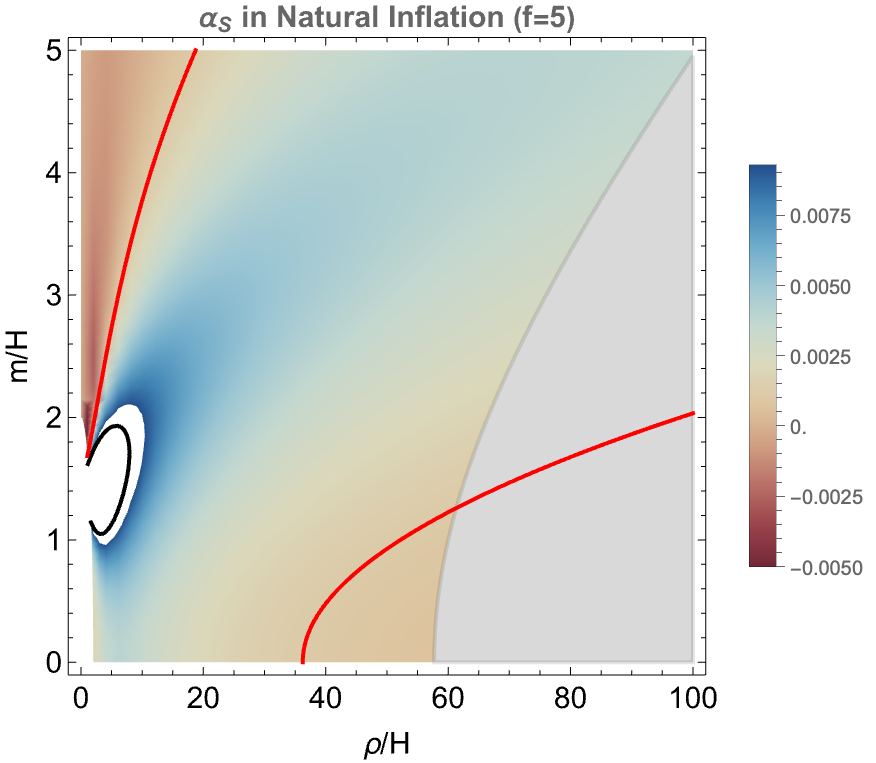}
  \end{center}
 \end{minipage}
\caption{{\it Left}: Contour plots of $n_s$ for $N=60$ as functions of $m/H$ and $\rho/H$ in Starobinsky inflation, chaotic inflation with $n=2$, and natural inflation with the axion decay constant $f=5$.  
The black (red) curve indicates the maximum (minimum) values of $n_s$ and $\alpha_s$ from Eq.~\eqref{ACT}.  
The region violating the EFT validity condition $m^2+\rho^2<4H^2$ is excluded from the figures. The gray region shows the bound from non-Gaussianity.   {\it Right}: Contour plots of $\alpha_s$ with the same parameter variations as in the left panel.
}
  \label{fig5}
\end{figure}

We now examine the effects of an additional massive field, or equivalently a modified sound speed with $c_s < 1$. As mentioned above, the general analytic expressions for the inflationary observables with $c_s$ are not particularly illuminating. Instead, we demonstrate the impact numerically by varying $m$ and $\rho$, while keeping the SR parameters fixed to the values given in Table~\ref{tab:model}. Fig.~\ref{fig1} and the upper panels of Fig.~\ref{fig5} show the results for Starobinsky inflation with $\alpha = 1$.  
Figures~\ref{fig2},~\ref{fig3}  and the middle panels of Fig.~\ref{fig5} correspond to chaotic inflation,  
while figures~\ref{fig4} and the lower panels of Fig.~\ref{fig5} illustrate the case of natural inflation with $f=5$. Further details are provided in the figure captions.

As shown, for each model (except for chaotic inflation with $n=4$) in Figs.~\ref{fig1}-~\ref{fig4}, there exists a parameter space that fits the ACT data by increasing both $n_s$ and $\alpha_s$.  
In particular, we find that a nonzero mixing $\rho$ plays an important role in realizing this.  
For comparison, in each plot, the prediction of the original models without an additional heavy field can be read off from the edge at $\rho=0$ of the heaviest mass (orange line).

Figure~\ref{fig5} shows the contour plots of $n_s$ and $\alpha_s$ as functions of $m/H$ and $\rho/H$ for each model, together with the observationally favored region at $1\sigma$, ~\eqref{ACT}.  
The region enclosed by the black and red curves satisfies the bounds given in Eq.~\eqref{ACT}.

\section{Non-Gaussian Features from Heavy Field}\label{sec:NG}
As we saw in the previous section, the presence of an additional heavy field with a mass of order the Hubble scale, together with non-vanishing mixing coupling, modifies the predictions for $n_s$ and $\alpha_s$ significantly through changes in the sound speed. In particular, we found that several models can be reconciled with the ACT results.

In this section, we show that such an additional field enhances non-Gaussianity compared to the single-field case~\cite{Maldacena:2002vr}, and even leaves specific imprints on it. This implies that the scenario can be observationally tested. In particular, we focus on the three-point correlation function of the curvature perturbation $\zeta$ (the bispectrum).

To discuss the bispectrum, we need to go beyond linear theory of perturbations. The cubic-order action of the curvature perturbation~$\zeta$ and the isocurvature perturbation~$\sigma$ in general two-field non-linear sigma models~\eqref{L_2_2} can be found in Ref.~\cite{Garcia-Saenz:2019njm, Pinol:2020kvw}. Among these, the dominant interactions contributing to the bispectrum through $\sigma$-exchange processes are classified in Ref.~\cite{Pinol:2021aun}. For example, cubic interactions involving two $\zeta$ fields and one $\sigma$ field are dominated by
\begin{align}
S=\int d\tau d^3x\  \frac{a^2\rho \dot{\phi}_0}{2H^2}\left[\left(\zeta^{\prime}\right)^2-\left(\partial_i \zeta\right)^2\right] \sigma,  \label{zeta^2sigma}
\end{align}
and, together with the linear mixing term proportional to $\eta_\perp$ (or $\rho$) in Eq.~\eqref{S_quad_a}, this contributes to the bispectrum through a single $\sigma$-exchange at tree level. Although there are also multiple $\sigma$-exchange processes~\cite{Pinol:2021aun, Xianyu:2023ytd, Aoki:2024uyi}, we focus on the interactions described above in this work for simplicity.

\subsection{Bispectrum in EFT}
First, let us compute the bispectrum within the single-field EFT, following the same approach as in the previous section. This method is often sufficient to estimate the size of bispectrum in the equilateral configuration when $\sigma$ is relatively heavy compared to the Hubble scale. After integrating out $\sigma$ using Eq.~\eqref{sigma_int}, we obtain derivative self-interaction terms for $\zeta$:\footnote{Other interactions such as $\dot{\zeta}\sigma^2$ and $\sigma^3$ which exist in the expansion of Eq.~\eqref{L_2_2}, also contribute to the effective $\dot{\zeta}^3$ term, as discussed in Ref.~\cite{Garcia-Saenz:2019njm}. 
Some of these contributions are comparable to Eq.~\eqref{int_a}, while others depend on model-specific details such as the target field-space metric and the potential form. 
These effects do not significantly alter the following estimation of non-Gaussianity in models where the contributions from the field-space geometry or the potential are subdominant, which we assume throughout this paper.}
\begin{align}
S=-\int d\tau d^3x \frac{a\epsilon \Mpl^2}{H}\frac{1-c_s^2}{c_s^2} \left[\left(\zeta^{\prime}\right)^2-\left(\partial_i \zeta\right)^2\right] \zeta^{\prime}+\cdots,\label{int_a}   
\end{align}
where, $\cdots$ denotes sub-leading terms, and the sound speed $c_s$ is understood as defined in Eq.~\eqref{c_s}. The bispectrum $\langle\zeta_{\mathbf{k}_1} \zeta_{\mathbf{k}_2} \zeta_{\mathbf{k}_3}\rangle$ arising from the contact interaction in Eq.~\eqref{int_a} can then be computed using the standard in-in formalism~\cite{Weinberg:2005vy, Chen:2017ryl, Wang:2013zva}. Defining the so-called shape function $S(k_1,k_2,k_3)$ as
\begin{align}
\langle\zeta_{\mathbf{k}_1} \zeta_{\mathbf{k}_2} \zeta_{\mathbf{k}_3}\rangle \equiv (2 \pi)^7 \frac{P_\zeta^2}{\left(k_1 k_2 k_3\right)^2} S(k_1,k_2,k_3)\, \delta^{(3)}(\mathbf{k}_1+\mathbf{k}_2+\mathbf{k}_3),
\end{align}
where $P_\zeta$ is the power spectrum defined in Eq.~\eqref{P_zeta}, we obtain
\begin{align}
\nonumber S(k_1,k_2,k_3)=&\ \frac{c_s^2-1}{16c_s^2} \frac{k_3}{k_1k_2(k_1 + k_2 + k_3)^{3}}\left[
  2 k_1^{4}
+ 3 k_1^{3} (2 k_2 + k_3)\right.\\
\nonumber  &+ (k_2 + k_3)^{2}\,(2 k_2^{2} - k_2 k_3 - k_3^{2}) +k_1^{2}\!\left( 4(1 + c_s^{2}) k_2^{2} + 3 k_2 k_3 - k_3^{2} \right)
\\
&\left.+ 3 k_1 \left( 2 k_2^{3} + k_2^{2} k_3 - 2 k_2 k_3^{2} - k_3^{3} \right)
\right]+ {\rm 5\ perms.},
\end{align}
where ${\rm 5\ perms.}$ denotes the remaining permutations of $k_{1,2,3}$. The result above is consistent with Ref.~\cite{Creminelli:2003iq}.  

We note that the shape function depends on model-specific details, such as the inflaton scalar potential, only implicitly through $c_s$ (or equivalently the mass $m$ and mixing coupling $\rho$ of $\sigma$). This allows us to compute it without specifying a particular model. 
For convenience, we further define the equilateral nonlinearity parameter
\begin{align}
 f_{\rm NL}^{\rm eq} \equiv \frac{10}{9} S(k,k,k)=\frac{5(c_s^2-1)(4c_s^2+17)}{324c_s^2}, \label{f_nl_eq}
\end{align}
which depends only on $m$ and $\rho$ through $c_s$, by virtue of scale invariance.
\begin{figure}[H]
    \centering
    \includegraphics[width=75mm]{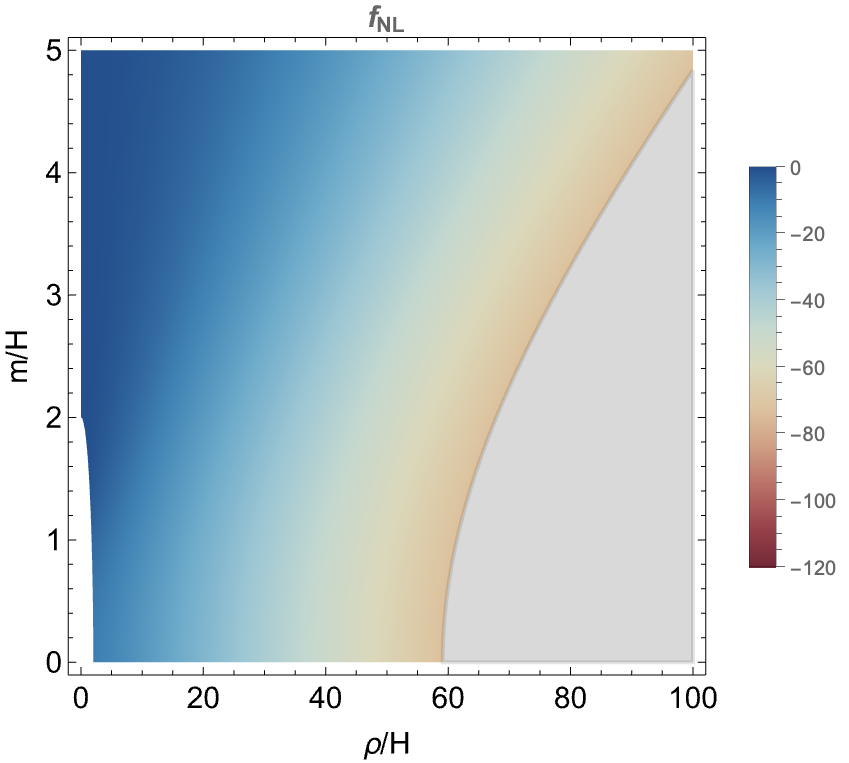}
    \caption{Contour plots of $f_{\rm NL}^{\rm eq}$ in Eq.~\eqref{f_nl_eq} as function of $m/H$ and $\rho/H$. The region violating the EFT validity condition $m^2+\rho^2<4H^2$ is excluded from the figures. The gray shaded region is excluded by Planck~\cite{Planck:2019kim}.}
    \label{fig6}
\end{figure}
In Fig.~\ref{fig6}, we show $f_{\rm NL}^{\rm eq}$ as functions of $m$ and $\rho$.  
As seen there, the region with large mixing predicts a large non-Gaussianity, and some parts are already excluded.  
Figure~\ref{fig7} shows the $(n_s,f_{\rm NL}^{\rm eq})$ plane with $\rho$ varied.  
As $\rho$ increases, some of the model predictions enter the ACT favored region~\eqref{ACT}, but some parameter space with large $\rho$ is excluded by the non-Gaussianity constraints~$f_{\mathrm{NL}}^{\mathrm{eq}}=-26\pm 47$~\cite{Planck:2019kim}.  
Even in the unconstrained region, the parameter space with large $\rho$ has the potential to be tested by future observations.

\begin{figure}[H]
 \begin{minipage}{0.5\hsize}
  \begin{center}
   \includegraphics[width=80mm]{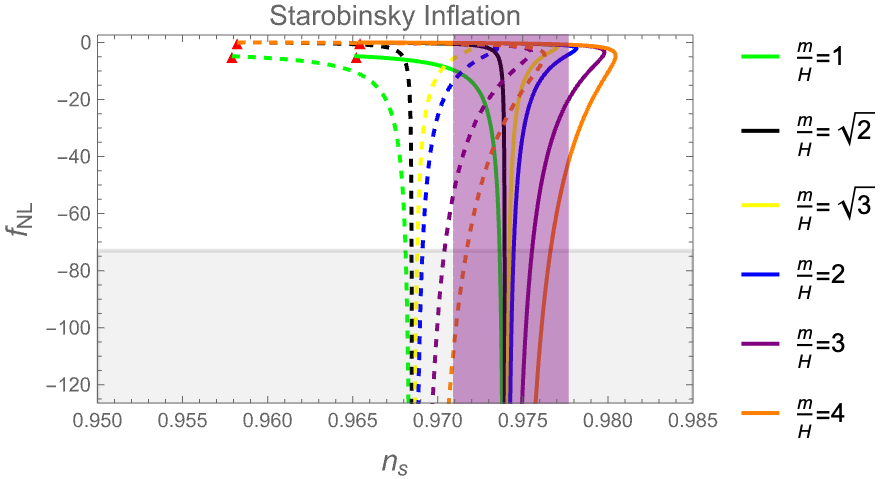}
  \end{center}
 \end{minipage}
  \begin{minipage}{0.5\hsize}
  \begin{center}
   \includegraphics[width=80mm]{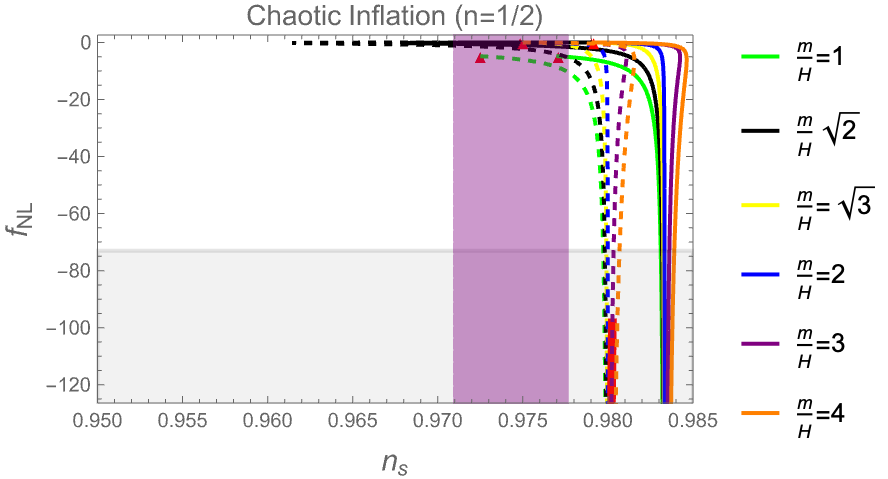}
  \end{center}
 \end{minipage}
 \begin{minipage}{0.5\hsize}
  \begin{center}
   \includegraphics[width=80mm]{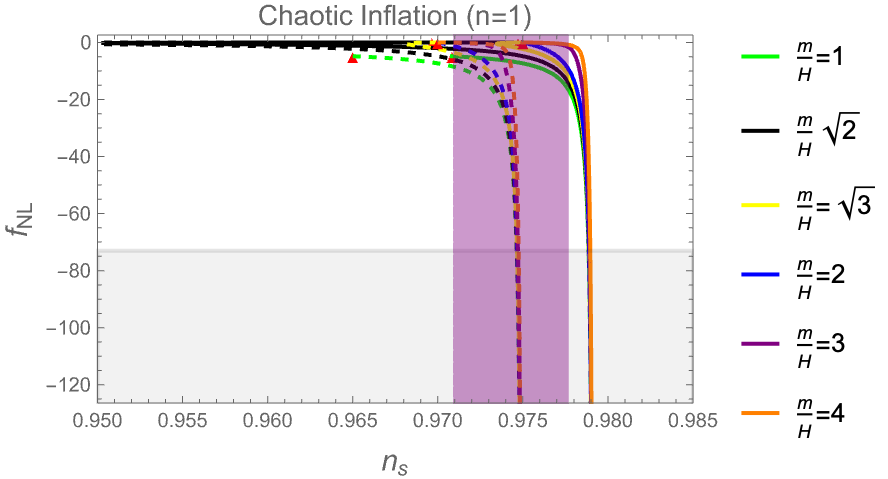}
  \end{center}
 \end{minipage}
 \begin{minipage}{0.5\hsize}
  \begin{center}
   \includegraphics[width=80mm]{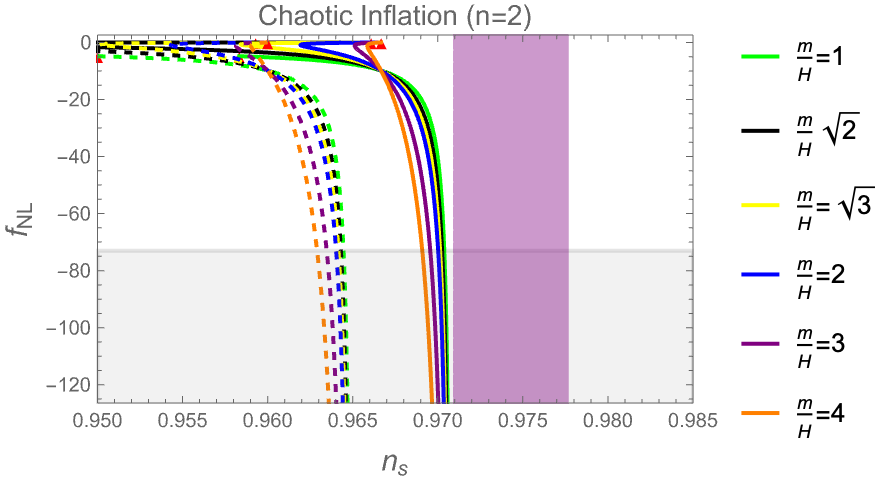}
  \end{center}
 \end{minipage}
 \begin{minipage}{0.5\hsize}
  \begin{center}
   \includegraphics[width=80mm]{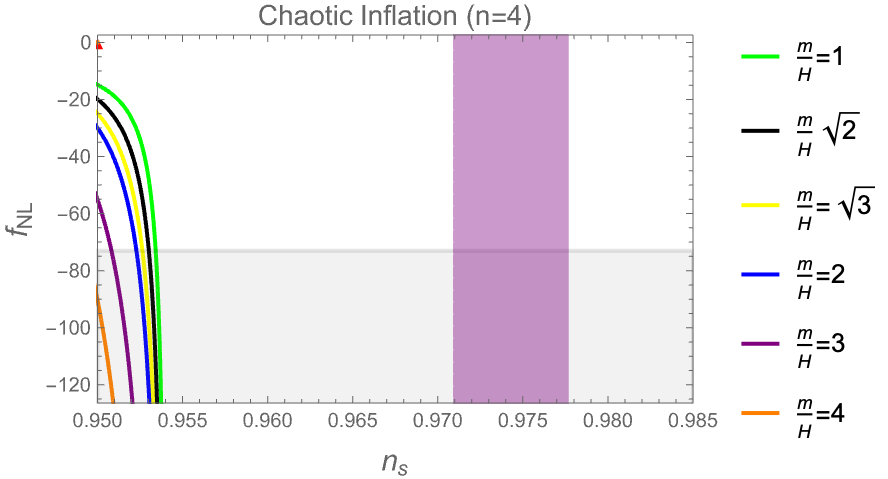}
  \end{center}
 \end{minipage}
 \begin{minipage}{0.5\hsize}
  \begin{center}
   \includegraphics[width=80mm]{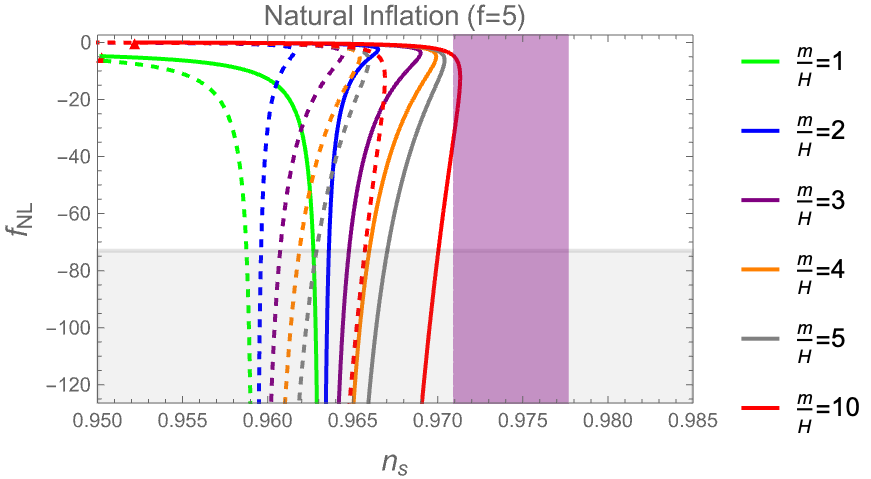}
  \end{center}
 \end{minipage}
\caption{$(n_s,f_{\rm NL}^{\rm eq})$ plane obtained by varying $\rho$ while fixing $m$ as indicated in each panel.  
Solid (dashed) curves correspond to $N=60$ ($N=50$).  
From top to bottom, we show examples of Starobinsky inflation, chaotic inflation with $n=1/2,1,2,4$, and natural inflation models.  
The dark purple region denotes the ACT favored region~\eqref{ACT} at $1\sigma$, while the gray shaded region is excluded by the Planck constraint~\cite{Planck:2019kim}.
}
  \label{fig7}
\end{figure}

\subsection{Cosmological collider signal}

The single-field EFT approach described above successfully captures both the shape and the magnitude of the bispectrum around the equilateral configuration. However, it misses an essential contribution from the particle production of $\sigma$, known as the cosmological collider signal~\cite{Chen:2009zp, Baumann:2011nk, Noumi:2012vr, Arkani-Hamed:2015bza}. This effect becomes particularly important when the mass of $\sigma$ is comparable to, or a few times larger than, the Hubble scale. In this regime, characteristic oscillatory features, determined by both $m$ and $\rho$, emerge in the squeezed limit, corresponding to correlations between long and short wavelength modes.

For small mixing, $\rho/H \ll 1$, the perturbative Feynman diagrammatic approach is applicable~\cite{Chen:2009zp}. For instance, the analytic expression of the single-$\sigma$ exchange contribution to the bispectrum arising from Eq.~\eqref{zeta^2sigma} is provided in~\cite{Arkani-Hamed:2018kmz, Qin:2023ejc}. In our case, however, we are interested in a broader parameter space, including the non-perturbative regime $\rho/H \gg 1$, where no analytic result is available. In this regime, one can instead rely on the numerical approach developed in~\cite{Werth:2023pfl, Pinol:2023oux, Werth:2024aui}, dubbed \textit{CosmoFlow}, which enables the estimation of both the shape and the amplitude of the bispectrum for arbitrary momentum configurations. Pursuing this direction is not the main focus of the present work. Nevertheless, as shown in~\cite{Werth:2023pfl}, oscillations with frequency 
\begin{align}
\sqrt{\tilde{m}^2+\tilde{\rho}^2-\tfrac{9}{4}},
\end{align}
should generically appear in the squeezed limit, carrying robust information of the massive field $\sigma$.

\section{Concrete Models}\label{sec:RTI}
As we saw in the previous section, a large mixing or a rapid turn plays a crucial role in modifying the inflationary predictions of single-field models. Up to this point, we have adopted an EFT perspective that remains agnostic about the details of the UV completion. In what follows, we present concrete realizations of such scenarios, where the inflaton couples to a heavy field with a Hubble-scale mass and follows a turning trajectory, thereby realizing several of the potential forms discussed in Section~\ref{sec:inflation models}.

\subsection{Axionic chaotic inflation}
As a simple example, we consider the following two-field system with flat target space described by radial (saxion) $r$ and angle (axion) $\theta$, 
\begin{align}
&S=\int \mathrm{d}^4 x \sqrt{-g}\left[\frac{1}{2}(\partial r)^2+\frac{1}{2} r^2(\partial \theta)^2-V(r, \theta)\right],\label{S_r_s}\\
&V(r, \theta)=\alpha \theta^n+\frac{M^2}{2}\left(r-R\right)^2,\label{V_r_s}
\end{align}
where $\alpha,n, M,$ and $R$ are real positive parameters.\footnote{A similar setup was studied in Ref.~\cite{Achucarro:2012yr}, where the potential takes the form 
\[
V = V_0 + \alpha \theta + \frac{M^2}{2}\left(r - R\right)^2 .
\]
Our construction is motivated by this, but differs in an important way. In Ref.~\cite{Achucarro:2012yr}, the constant term $V_0$ dominates during inflation so that $H \simeq \sqrt{V_0/3}$, and $r$ remains almost exactly constant. By contrast, in our setup the radial direction $r$ acquires a mild $\theta$-dependence, $r = r(\theta)$, as we will see below. A more general discussion of the $(r,\theta)$ system with a non-trivial field-space metric can be found in Ref.~\cite{Aragam:2021scu}, although some concrete models analyzed there differ from ours. } This kind of potential for the axion $\theta$ could arise from string compactifications, for example through the dimensional reduction of higher-form fields in the Dirac-Born-Infeld action, as in axion monodromy inflation~\cite{McAllister:2008hb, Silverstein:2008sg}. The radial direction corresponds to the modulus field controlling the volume of the internal cycle in the extra-dimensional space.

The background equations of motion for $r(t)$ and $\theta(t)$ in the FLRW background are given by
\begin{align}
&\ddot{\theta}+3 H \dot{\theta}+2 \dot{\theta} \frac{\dot{r}}{r}=-n\alpha \frac{\theta^{n-1}}{r^2},\label{e1}\\
&\ddot{r}+3 H \dot{r}+r\left(M^2-\dot{\theta}^2\right)= M^2R,\label{e2}
\end{align}
where the dot denotes the time derivative.
Now let us consider the situation where $\dot{r}, \ddot{r},$ and $\ddot{\theta}$ are negligible, which means that $\theta$-direction enjoys the usual SR dynamics while the $r$-direction is nearly constant. In this sense, the angular (axion) direction plays the role of the inflaton, and we refer to this scenario as {\it{axionic chaotic inflation}}, reflecting the potential form in Eq.~\eqref{V_r_s} with $r \sim \mathrm{const.}$.
Then, EOMs above are reduced to 
\begin{align}
3 H \dot{\theta}\simeq -n\alpha \frac{\theta^{n-1}}{r^2}, \quad \dot{\theta}^2\simeq M^2\left(1-\frac{R}{r}\right), \label{SR_AS}
\end{align}
which are supplemented by the Friedmann equation with the same SR conditions:
\begin{align}
 3H^2\simeq \frac{1}{2}r^2\dot{\theta}^2+V.    
\end{align}
The system is closed in the sense that one can solve $r,\dot{\theta},$ and $H$ as functions of $\theta$.

We further assume  
\begin{align}
\frac{R}{r} \ll 1,\label{assmp}   
\end{align}
under which Eq.~\eqref{SR_AS} reduces to
\begin{align}
\frac{n^2\alpha ^2 \theta^{2 n-2}}{r^4}\simeq 3 m^2 \left(m^2 r^2+\alpha  \theta^n\right),
\end{align}
from which we obtain an approximate analytic solution for $r$: 
\begin{align}
r^2(\theta)\simeq&\ 
\frac{\alpha  \theta^n}{3 M^2}\left[\left(\gamma(\theta)^{1/3}+\gamma(\theta)^{-1/3}\right)-1\right], \quad \dot{\theta}\simeq  -M, \label{sol_1}
\end{align}
with
\begin{align}
\gamma(\theta)\equiv \frac{3M n}{2\alpha\theta^{n+2}} \left(3 M n+\sqrt{9 M^2 n^2-4 \alpha \theta^{n+2}}\right)-1. 
\end{align}
The SR parameters $\epsilon$ and $\eta$ are estimated as 
\begin{align}
 \epsilon \equiv -\frac{\dot{H}}{H^2} \simeq \frac{3 M^2 r^2}{2 \left(\alpha  \theta^n+M^2 r^2\right)},\quad
 \eta \equiv \frac{\dot{\epsilon}}{\epsilon H}\simeq \frac{\sqrt{3} M n \alpha \theta^{n-1}}{\left(\alpha  \theta^n+M^2 r^2\right)^{3/2}},\label{epsilon_a}
\end{align}
where $r$ is understood as Eq.~\eqref{sol_1}.

Based on the analytic solution in Eq.~\eqref{sol_1}, the turn rate $\eta_\perp$ (see Eq.~\eqref{turn_2} for the explicit expression), which is related to the mixing coupling through $\rho = 2 \eta_\perp H$, is evaluated as
\begin{align}
\eta_\perp H = -M \ \ (= \dot{\theta}) \, ,
\end{align}
which can be rewritten using Eqs.~\eqref{sol_1} and~\eqref{epsilon_a} as
\begin{align}
\label{eq:Turn_anal}
\eta_\perp^2 \simeq \frac{2\epsilon}{r^2} \, .
\end{align}
Note that this is consistent with Eq.~\eqref{eta_perp_a} with $\kappa=1/r$. Also, it implies that in order to realize a large turn (or equivalently, a large mixing coupling), we need to require a large curvature,
$r \sim \sqrt{\epsilon}$. Such a situation is achieved when the hierarchy $\alpha/M^2 \ll 1$ is satisfied. Finally, the isocurvature mass $m$ (defined in Eq.~\eqref{m_sigma}), in our case, takes the form
\begin{align}
\label{eq:miso_anal}
m^2 = \frac{M^2 R}{r} \, .
\end{align}

We have numerically verified that the inflationary solution~\eqref{sol_1} can indeed be realized. In Table~\ref{tab:turn}, we present several benchmark examples. The parameters are chosen such that they satisfy the condition~\eqref{assmp} and reproduce the observed CMB normalization, $P_\zeta = 2.1 \times 10^{-9}$. For each example, we also list the resulting inflationary observables\footnote{Note that $r$ refers to the tensor-to-scalar ratio, not the saxion field value.} together with the turn rate, $\eta_\perp \equiv \rho/(2H)$, and the mass of the isocurvature mode, $m$, evaluated at $N=60$ (corresponding to the CMB pivot scale). We find that all examples realize large mixing $\rho/H\gtrsim 1$ and isocurvature mass around the Hubble scale, which corresponds to the cases we discussed in the section~\ref{sec:EFT}.

\begin{figure}[H]
 \begin{minipage}{0.5\hsize}
  \begin{center}
   \includegraphics[width=72mm]{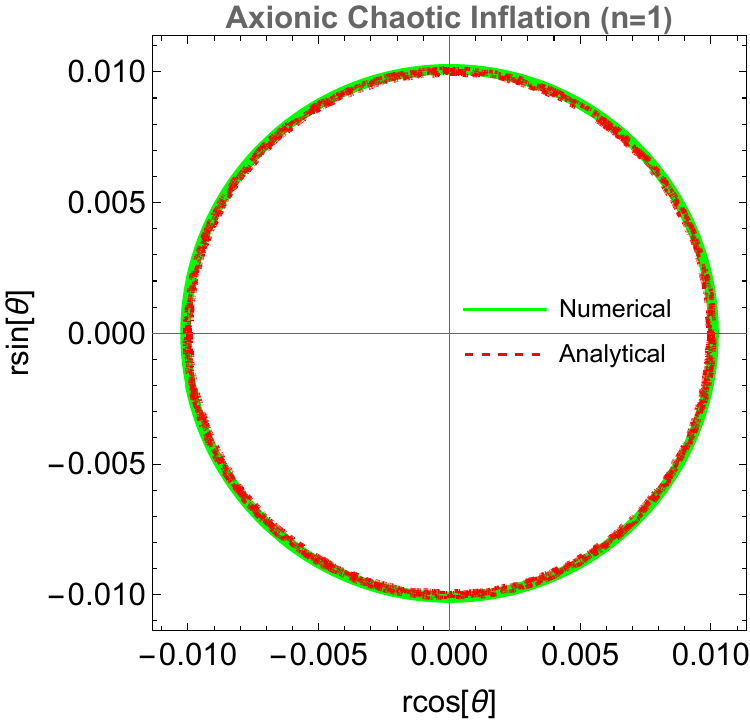}
  \end{center}
 \end{minipage}
 \begin{minipage}{0.5\hsize}
  \begin{center}
   \includegraphics[width=72mm]{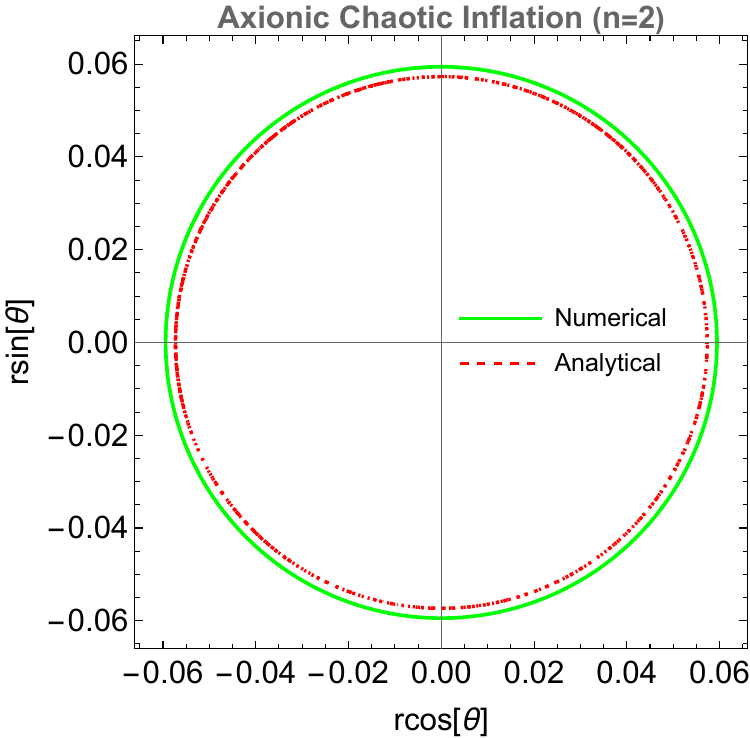}
  \end{center}
 \end{minipage}
 \begin{minipage}{0.5\hsize}
  \begin{center}
   \includegraphics[width=75mm]{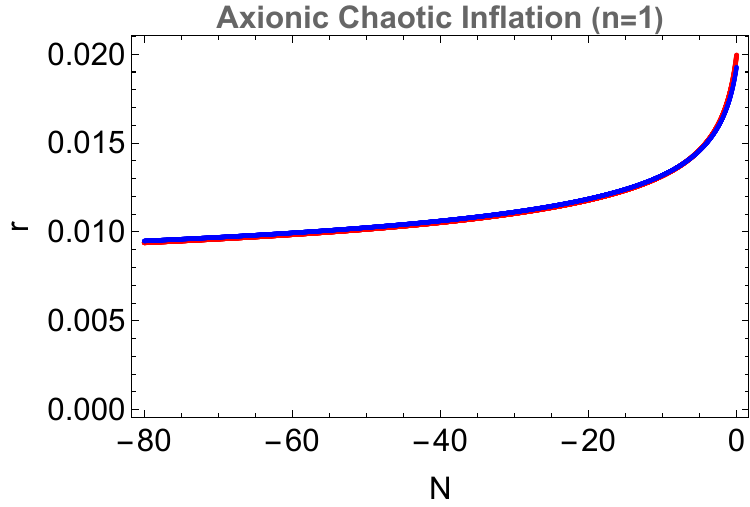}
  \end{center}
 \end{minipage}
 \begin{minipage}{0.5\hsize}
  \begin{center}
   \includegraphics[width=75mm]{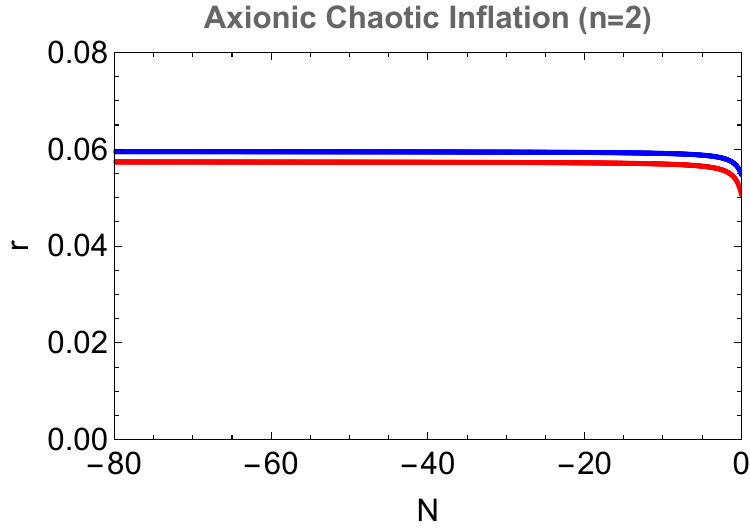}
  \end{center}
 \end{minipage}
  \begin{minipage}{0.5\hsize}
  \begin{center}
   \includegraphics[width=75mm]{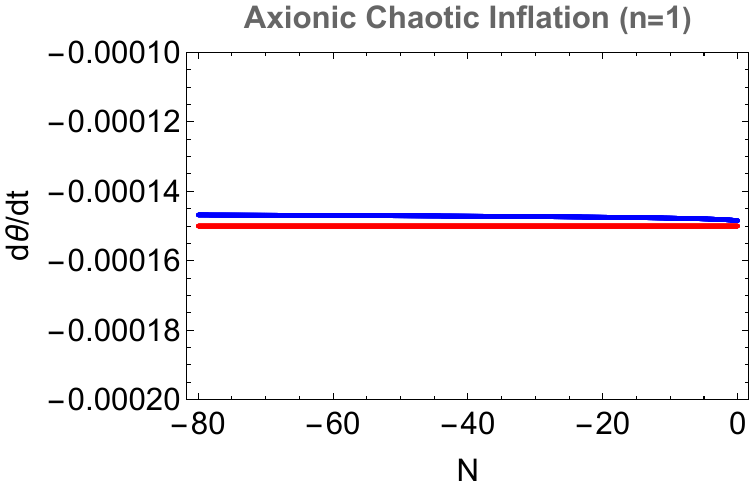}
  \end{center}
 \end{minipage}
 \begin{minipage}{0.5\hsize}
  \begin{center}
   \includegraphics[width=75mm]{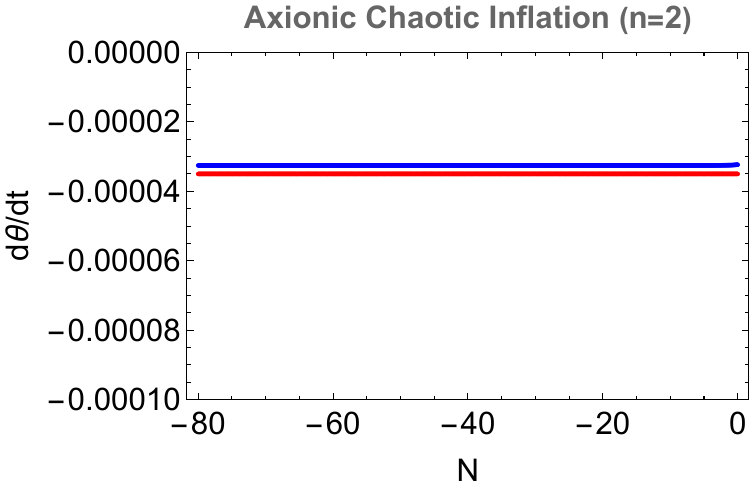}
  \end{center}
 \end{minipage}
    \caption{Inflationary trajectories in polar coordinates $(r,\theta)$ and the time evolution of $r$ and $\dot{\theta}$, from top to bottom. 
The left (right) panels correspond to $n=1$ ($n=2$). 
In the top panels, the numerical solutions of the background equations~\eqref{e1} and~\eqref{e2} are shown in green, while the analytical approximation $r=r(\theta)$ in Eq.~\eqref{sol_1} is shown in red. Here we show the results for the range between the last $60$ and $50$ ($30$) e-folds before the end of inflation for $n=1(2)$. In the middle panels, the blue curves represent the numerical solutions $r(N)$ for the last $80$ e-folds, while the red curves denote the semi-analytical solutions: $r(\theta)$ from Eq.~\eqref{sol_1} with the numerically estimated $\theta(N)$ inserted.  
In the bottom panels, the blue curves represent the numerical solutions $\dot{\theta}(N)$, while the red curves show the analytical solution $\dot{\theta}=-M$.}
    \label{fig:trajofr_p1}
\end{figure}

\begin{table}[H]
\centering
\caption{Benchmark examples for axionic chaotic inflation.}
\label{tab:turn}
\renewcommand{\arraystretch}{1.5} 
\begin{tabular}{ccccc}
\toprule
\textbf{Parameters:} \((n,\alpha,M,R)\) & \( n_s \) & \( r \) & \( |\eta_\perp|\equiv |\rho|/(2H) \) & \( m/H \)  \\
\midrule
\(\left(\frac{1}{2},2.7\times 10^{-11},3 \times 10^{-4}, 4 \times 10^{-4}\right)\) & \( 0.98 \)& \(0.03\) &\(16.2\) &\( 4.77  \)\\

\(\left(1,6.1\times 10^{-13},1.5 \times 10^{-4}, 4 \times 10^{-4}\right)\) & \( 0.975 \)&\(0.018\)&\(10.5\)&\(2.15\)\\   

\(\left(2,1\times 10^{-14},3.5 \times 10^{-5}, 8 \times 10^{-4}\right)\) & \( 0.967 \)&\(0.022\)&\(2.16\)&\(0.85\)\\
\bottomrule
\end{tabular}
\end{table}

\begin{figure}[H]
 \begin{minipage}{0.5\hsize}
  \begin{center}
   \includegraphics[width=75mm]{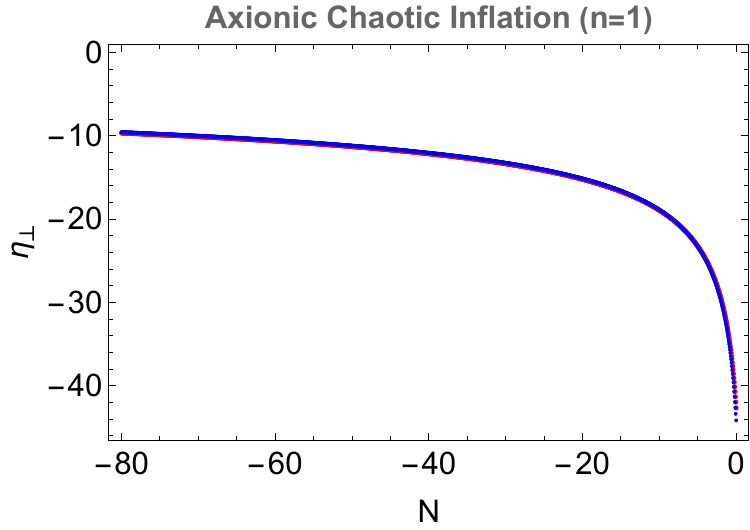}
  \end{center}
 \end{minipage}
 \begin{minipage}{0.5\hsize}
  \begin{center}
   \includegraphics[width=75mm]{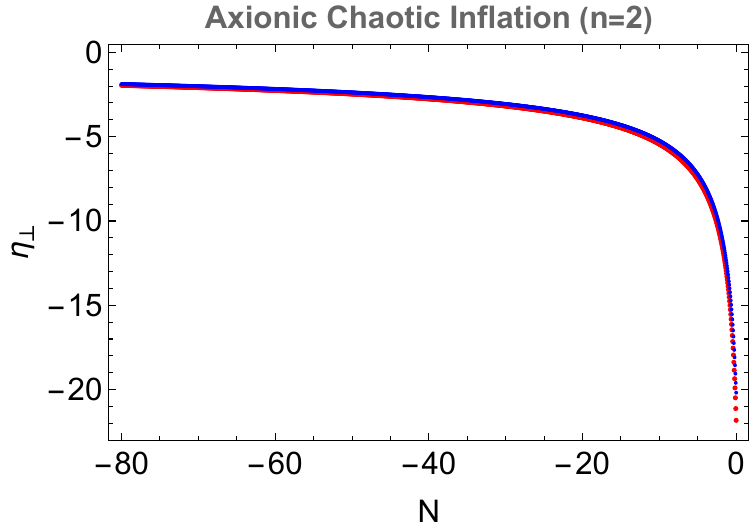}
  \end{center}
 \end{minipage}
 \begin{minipage}{0.5\hsize}
  \begin{center}
   \includegraphics[width=75mm]{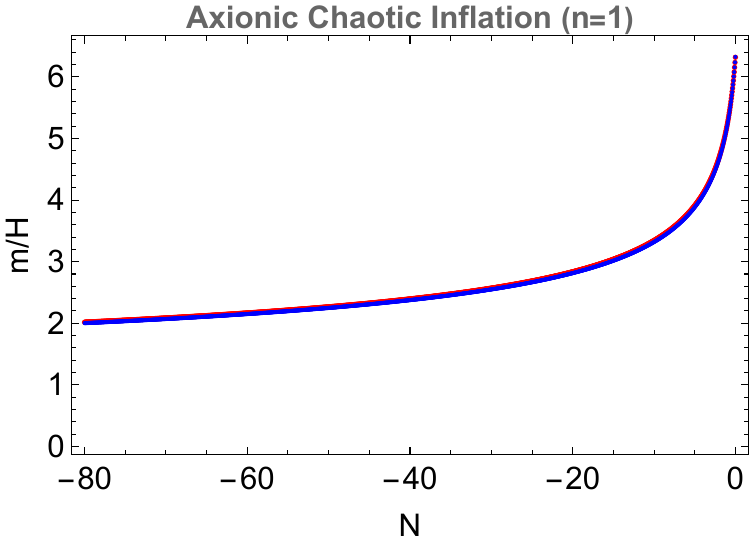}
  \end{center}
 \end{minipage}
 \begin{minipage}{0.5\hsize}
  \begin{center}
   \includegraphics[width=75mm]{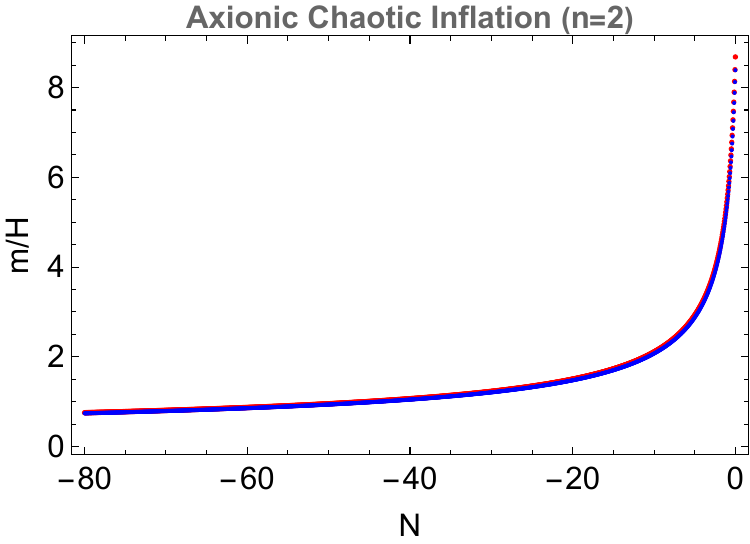}
  \end{center}
 \end{minipage}
     \caption{Time evolution of the turn rate $\eta_\perp$ and the isocurvature mass $m$. 
The full numerical solutions are shown by the blue curve, while the analytical expressions~\eqref{eq:Turn_anal} and~\eqref{eq:miso_anal} are shown by the red curve.
}
    \label{fig:trajofr_p2}
\end{figure}

\begin{figure}[H]
 \begin{minipage}{0.5\hsize}
  \begin{center}
   \includegraphics[width=75mm]{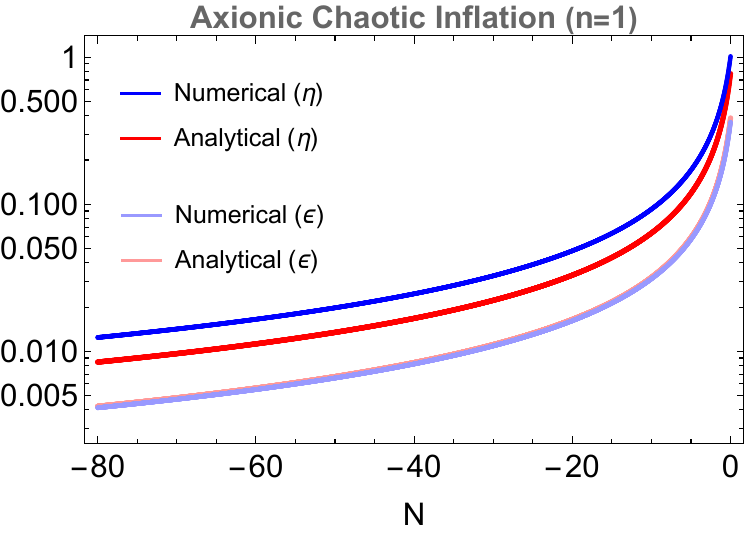}
  \end{center}
 \end{minipage}
 \begin{minipage}{0.5\hsize}
  \begin{center}
   \includegraphics[width=75mm]{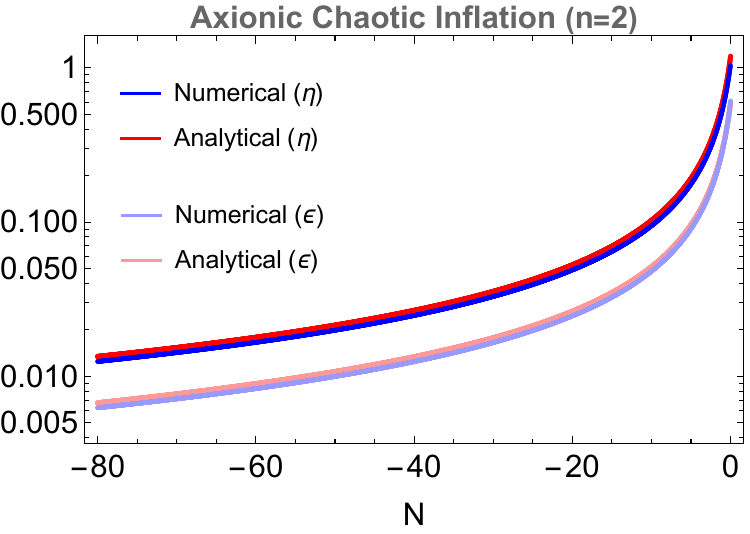}
  \end{center}
 \end{minipage}
     \caption{Time evolution of the slow-roll parameters $\epsilon$ and $\eta$. 
The full numerical solutions are shown by the blue curve, while the analytical ones~\eqref{epsilon_a} are shown by the red curve.
}
    \label{fig:trajofr_p3}
\end{figure}

Figure~\ref{fig:trajofr_p1} shows the inflationary trajectory and the time evolution of $r$ and $\dot{\theta}$, while Fig.~\ref{fig:trajofr_p2} presents the time evolution of the turn rate and the isocurvature mass.  
We find that the analytic solution~\eqref{sol_1}, together with Eqs.~\eqref{eq:Turn_anal} and~\eqref{eq:miso_anal}, capture numerical solutions very well over most of the e-fold range. In addition, the slow-roll parameters have the same behavior between the analytic solutions~\eqref{epsilon_a} and numerical ones, as shown in Fig.~\ref{fig:trajofr_p3}.
Further details are provided in the figure caption.  
We have also confirmed that the same holds for the $n=1/2$ case.

\subsection{Axionic Starobinsky inflation}
In the same spirit as in the previous subsection, we attempt to implement Starobinsky-type inflation in the axion-saxion system of Eq.~\eqref{S_r_s}. We choose the scalar potential as
\begin{align}
V(r, \theta)=V_0\left( 1 - e^{-\sqrt{\frac{2}{3}}\,\theta}\right)^2+\frac{M^2}{2}\left(r-R\right)^2,
\end{align}
where $V_0, M, R$ are real parameters.\footnote{The same terminology is used in Ref.~\cite{Blumenhagen:2015qda}, where the interesting possibility of realizing Starobinsky-type inflation in string theory was considered. However, this corresponds to a different model.}

The SR equations corresponding to Eq.~\eqref{SR_AS} are now given by 
\begin{align}
3 H \dot{\theta}\simeq -2\sqrt{\tfrac{2}{3}}\frac{V_0 e^{-2\sqrt{\tfrac{2}{3}}\theta }}{r^2}, 
\qquad 
\dot{\theta}^2\simeq M^2\left(1-\frac{R}{r}\right), 
\label{SR_AS_ST}
\end{align}
for $\theta \gg 1$.\footnote{The opposite limit $\theta \ll 1$ reduces to the chaotic inflation case with $n=1$, which we have discussed in the previous subsection. We therefore do not consider this situation here.} 
In this regime, however, it is difficult to satisfy the ``rapid turn condition'' $R/r \ll 1$ because of the exponential suppression of $r$. More explicitly, the solutions of Eq.~\eqref{SR_AS_ST} obtained under the assumption $R/r \ll 1$ are 
\begin{align}
r^2(\theta)\simeq \frac{2\sqrt{2}\,V_0^{1/2}e^{- \sqrt{\frac{2}{3}} \theta}}{3m},
\qquad 
\dot{\theta}\simeq  -M,
\end{align}
from which one can see that $r\ll 1$ for $\theta \gg 1$.

Here, instead, we seek solutions under the ``slow turn condition''\footnote{This definition is meant only for comparison with the axionic chaotic inflation considered in the previous subsection. In the literature~\cite{Bjorkmo:2019fls}, inflation models satisfying $\eta^2_\perp \gg \mathcal{O}(\epsilon)$ are classified as rapid-turn inflation models. The axionic Starobinsky inflation studied here satisfies this criterion.} $r\simeq R$, namely 
\begin{align}
r=R+\delta r, \qquad \text{with} \quad \delta r / R \ll 1 .
\label{eq:rStaro}
\end{align}
In this case, we find another type of solutions to Eq.~\eqref{SR_AS_ST} as
\begin{align}
\delta r\simeq \frac{8 V_0}{9 m^2 R^3} e^{-2 \sqrt{\tfrac{2}{3}} \theta}, 
\qquad 
\dot{\theta} \simeq -\frac{2\sqrt{2}\, V_0^{1/2}}{3 R^2} e^{-\sqrt{\tfrac{2}{3}} \theta}.\label{app_slow}
\end{align}
From these, the turn rate $\eta_\perp$ and the isocurvature mass $m$ are estimated as
\begin{align}
\eta_{\perp} \simeq -2 \sqrt{\tfrac{2}{3}} \frac{e^{-\sqrt{\tfrac{2}{3}}\theta}}{R^2}, 
\qquad 
m^2 \simeq M^2 . \label{app_slow2}
\end{align}
Since the SR parameters $\epsilon$ and $\eta$ are now estimated as 
\begin{align}
\epsilon \simeq \frac{4e^{-2\sqrt{\tfrac{2}{3}}\theta}}{3R^2},
\quad
\eta \simeq \frac{8\sqrt{2V_0}e^{-\sqrt{\tfrac{2}{3}}\theta}}{3R^2\sqrt{2V_0 + m^2R^2 }},
\end{align}
the turn rate can be written as 
\begin{align}
\eta_\perp^2\simeq \frac{2\epsilon}{R^2},    
\end{align}
which is consistent with Eq.~\eqref{eta_perp_a} with $\kappa=1/R$.

\begin{figure}[htbp]
  \centering
 
 \begin{minipage}{0.48\hsize}
  \begin{center}
   \includegraphics[width=70mm]{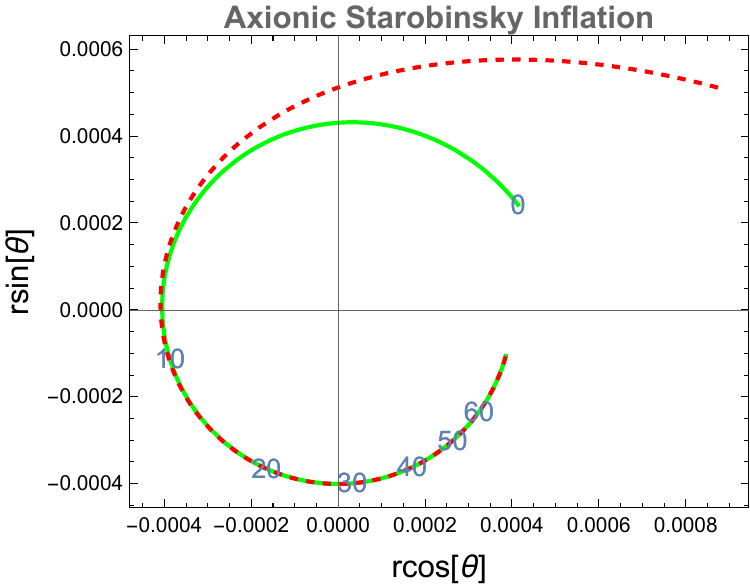}
  \end{center}
 \end{minipage}
 \begin{minipage}{0.48\hsize}
  \begin{center}
   \includegraphics[width=70mm]{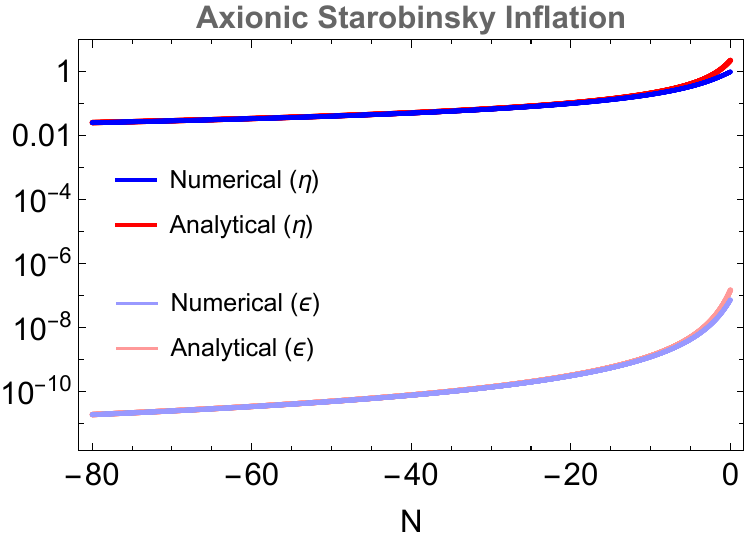}
  \end{center}
 \end{minipage}
      
  \vspace{0.5em}

 \begin{minipage}{0.48\hsize}
  \begin{center}
   \includegraphics[width=70mm]{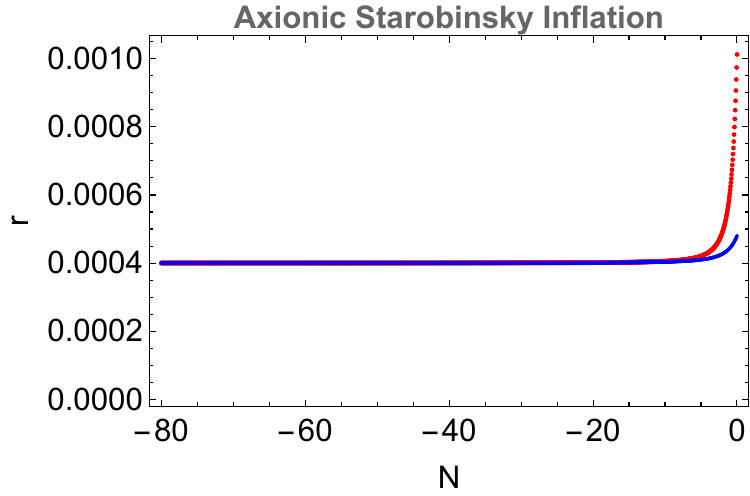}
  \end{center}
 \end{minipage}
 \begin{minipage}{0.48\hsize}
  \begin{center}
   \includegraphics[width=70mm]{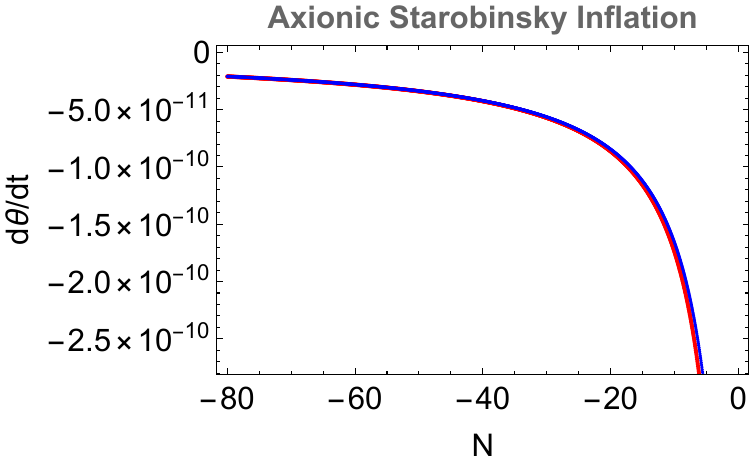}
  \end{center}
 \end{minipage}

  \caption{Same as Fig.~\ref{fig:trajofr_p1} and~\ref{fig:trajofr_p3}, but for axionic Starobinsky inflation. 
The corresponding number of e-folds from $N=0$ (inflation end) to $N=60$ is shown in the top left panel. Approximated analytic solutions (red line) are from Eq.~\eqref{app_slow}.}
 \label{fig:trajofr_Staro}
\end{figure}

\begin{figure}[htbp]
 \begin{minipage}{0.5\hsize}
  \begin{center}
   \includegraphics[width=75mm]{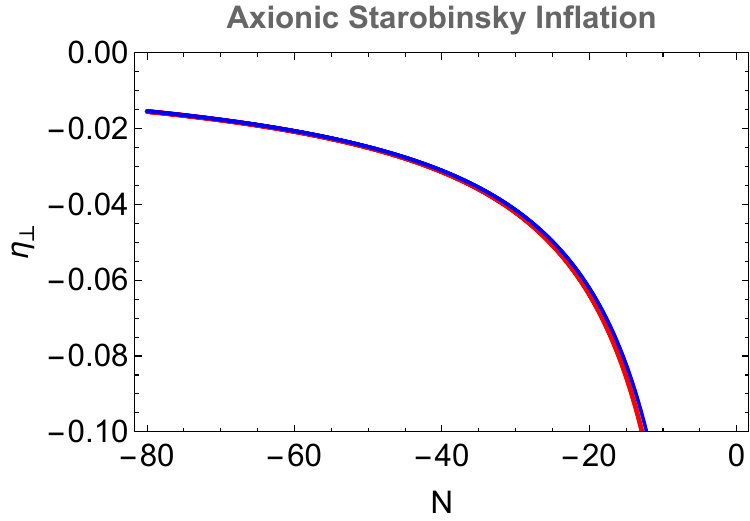}
  \end{center}
 \end{minipage}
 \begin{minipage}{0.5\hsize}
  \begin{center}
   \includegraphics[width=75mm]{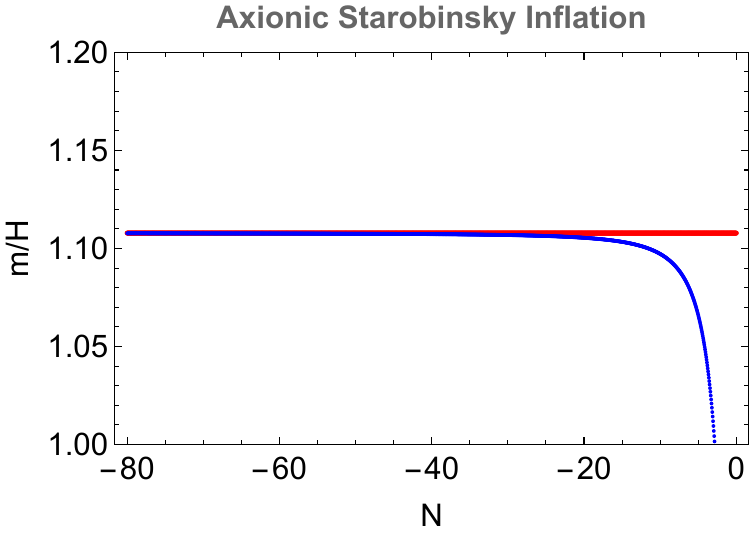}
  \end{center}
 \end{minipage}
     \caption{Same as Fig.~\ref{fig:trajofr_p2}, but for axionic Starobinsky inflation. Approximated analytic solutions (red line) are from Eq.~\eqref{app_slow2}. }
    \label{fig:trajofr_Staro2}
\end{figure}

We numerically checked that the approximate analytic solutions above provide a good description of the inflationary dynamics for most of the e-folds (see Figs.~\ref{fig:trajofr_Staro} and~\ref{fig:trajofr_Staro2}).
As a benchmark, we choose the parameter values listed in Table~\ref{tab:turn2} and present the corresponding inflationary observables, together with the turn rate and the isocurvature mass.
\begin{table}[h]
\centering
\caption{Benchmark example for axionic Starobinsky inflation.}
\label{tab:turn2}
\renewcommand{\arraystretch}{1.5} 
\begin{tabular}{ccccc}
\toprule
\textbf{Parameters:} \((V_0,M,R)\) & \( n_s \) & \( r \) & \( |\eta_\perp|\equiv |\rho|/(2H) \) & \( m/H \)  \\
\midrule
\(\left(5.5\times 10^{-18},1.5 \times 10^{-9}, 4 \times 10^{-4}\right)\) & \( 0.966 \)& \(1.78\times 10^{-10}\) &\(0.021\) &\( 1.11  \)\\
\bottomrule
\end{tabular}
\end{table}

\section{Summary and Discussion}\label{sec:summary}
In this paper, we examined the effects of an additional field with a Hubble-scale mass and a mixing coupling to the inflaton on inflationary observables ($n_s, r,$ and $\alpha_s$). We found that Starobinsky-type inflation models, whose predictions show a $2\sigma$ discrepancy with the recent ACT data, can be reconciled with the observations (and even chaotic inflation and natural inflation models, which are already excluded by Planck, can be revived), particularly in the case of large mixing couplings. Our scenario generally predicts large non-Gaussianity due to the strong mixing, which can be tested in future observations. We also discussed concrete realizations of models with a mixing coupling (turn), starting from a two-field setup, and presented explicit examples. In particular, we found that a rapid turn, $\rho/H \gtrsim 1$, can be achieved in chaotic-type inflation, while a mild turn, $\rho/H \gtrsim 10^{-2}$, is realized in Starobinsky-type inflation.

There are several promising directions for future research. 

First, extending our analysis to multiple fields with Hubble-scale masses is a natural but non-trivial generalization, as we have focused here on a single heavy field. 
Such setups with multiple Hubble-mass fields arise more naturally in supergravity and string-theoretic frameworks, and can lead to rich phenomenology for non-Gaussianity~\cite{Aoki:2020zbj, Pinol:2021aun}. 

Second, while we presented a simplified example of a large-turn two-field model, it would be interesting to embed such scenarios into UV-motivated frameworks. 
Our toy model~\eqref{S_r_s} represents one of the simplest realizations of a two-field system with a flat target space, effectively yielding single-field inflation with an additional Hubble-mass field and non-vanishing turn rate $\rho$.
In the literature, many examples of multifield inflation with (strongly) non-geodesic motion and (rapid) turns have been discussed, see e.g., Refs.~\cite{Brown:2017osf, Mizuno:2017idt, Bjorkmo:2019aev, Christodoulidis:2018qdw, Cremonini:2010ua, Renaux-Petel:2015mga, Renaux-Petel:2017dia, Garcia-Saenz:2018ifx, Grocholski:2019mot, Fumagalli:2019noh, Christodoulidis:2019jsx, Easson:2007dh, Achucarro:2010jv, Achucarro:2010da, Achucarro:2012sm, Achucarro:2012yr, Cespedes:2012hu, Hetz:2016ics, Chen:2018uul, Chen:2018brw, Achucarro:2019mea, Welling:2019bib, Achucarro:2019pux, Christodoulidis:2025wew}, mostly involving non-trivial field-space geometries (with the exception of Ref.~\cite{Achucarro:2012yr}). 
Some of these models exhibit attractor behavior, and the associated consistency conditions have been analyzed in Refs.~\cite{Bjorkmo:2019fls, Anguelova:2022foz, Anguelova:2024akm, Wolters:2024vzk}. 
Although embedding our setup into such scenarios is not straightforward, exploring possible connections would be an interesting direction for future work. 
We also note that realizing rapid turns within supergravity appears to face certain challenges~\cite{Aragam:2021scu}.

Third, in this paper, we do not consider the combined effects of classical excitations on top of quantum particle production for the minimal setup. 
Such combined effects can generally enhance non-Gaussianity or the cosmological collider signal through resonance~\cite{Chen:2008wn,Chen:2011zf,Chen:2014cwa,Braglia:2021ckn,Braglia:2021rej,Quintin:2024boj,Wang:2025qww} 
and additional scale dependence~\cite{Flauger:2016idt,Reece:2022soh,Aoki:2023wdc}. 
It would therefore be interesting to study these effects in our setup, 
especially in the context of the axionic inflation scenario discussed in Section~\ref{sec:RTI}. 
In particular, instanton effects can generically produce a cosine potential superimposed on the background potential discussed in this paper, 
which can act as a classical oscillation source and modify the predictions of inflationary observables.

We leave these extensions for future investigation.

\paragraph{Acknowledgments}
We would like to thank  Sergei V. Ketov for useful discussion. S.A. is supported by the Japan Science and Technology Agency (JST) as part of Adopting Sustainable Partnerships for Innovative Research Ecosystem (ASPIRE),  Grant No. JPMJAP2318. This work was supported in part by JSPS KAKENHI Grant Numbers JP25H01539 (H.O.).

\begin{appendix}

\section{Generalities for Two-Field Inflation}
\label{app:two}
Here we summarize some basics for two-field inflation (see Refs.~\cite{Gong:2016qmq, Wang:2013zva} for review).

We consider a system with two scalars $\phi^a=\{ \phi^1, \phi^2\}$
\begin{align}
S=\int d^4 x \sqrt{-g}\left[\frac{\Mpl^2}{2}R-\frac{1}{2} G_{ab} (\phi)\partial_\mu \phi^a \partial^\mu \phi^b-V\left(\phi\right)\right],   \label{L_2}
\end{align}
where $G_{ab} (\phi)$ and $V\left(\phi\right)$ are a field space metric and scalar potential. 

\subsection{Background evolution}
The background evolution of spacetime is described by the homogeneous and isotropic FLRW metric
\begin{align}
d s^2=-d t^2+a^2(t) d x^2,    
\end{align}
where $a(t)$ is the scale factor. The Hubble rate is defined as $H\equiv \dot{a}/a$, where the dot denotes a derivative with respect to cosmic time $t$, i.e., $\dot{a}=da/dt$. The background values of the scalar fields are denoted by $\phi_0^a(t)$.

The equations of motion for the background fields are given by
\begin{align}
&\mathcal{D}_t \dot{\phi}_0^a+3 H \dot{\phi}_0^a+G^{ab} V_b=0,\label{EOM1}\\
&3 \Mpl^2 H^2=\frac{1}{2} \dot{\phi}_0^2+V,\\
&\Mpl^2 \dot{H}=-\frac{1}{2} \dot{\phi}_0^2,\label{EOM3}
\end{align}
where the covariant derivative in field space is defined as
\begin{align}
\mathcal{D}_t A^a \equiv \dot{A}^a+\Gamma_{bc}^a \dot{\phi}_0^b A^c,    
\end{align}
with $\Gamma_{bc}^a$ the Christoffel symbols constructed from the field-space metric $G_{ab}$. The total field-space velocity is
\begin{align}
 \dot{\phi}_0^2\equiv G_{ab} \dot{\phi}_0^a \dot{\phi}_0^b, 
\end{align}
and $V_a=\partial V/\partial \phi^a$. Field-space indices $a$ are raised and lowered using $G_{ab}$ and its inverse $G^{ab}$. 

To describe the two-field dynamics, it is convenient to introduce the tangent vector $T^a$ and the normal vector $N_a$ to the field trajectory~\cite{Achucarro:2010da, Cespedes:2012hu},
\begin{align}
T^a=\frac{\dot{\phi}_0^a}{\dot{\phi}_0},\qquad N_a=(\operatorname{det} G)^{1 / 2} \epsilon_{ab} T^b,    \label{basis}
\end{align}
where $\epsilon_{12}=-\epsilon_{21}=1$, $\epsilon_{11}=\epsilon_{22}=0$.\footnote{This definition of $N_a$ is only applicable to a two-field system. More generally, one defines $N^a\propto \mathcal{D}_tT^a$~\cite{Achucarro:2010da}.} One can directly verify that $T^aT_a=N^aN_a=1$ and $T^aN_a=0$. 

The turning rate $\eta_{\perp}$ is then defined as 
\begin{align}
\eta_{\perp}=-\frac{N_a \mathcal{D}_t T^a}{H},    \label{turn_app}
\end{align}
which quantifies the deviation of the trajectory from a geodesic in field space. Projecting Eq.~\eqref{EOM1} along $T_a$ and $N_a$ yields
\begin{align}
\ddot{\phi}_0+3 H \dot{\phi}_0+V_T=0,\quad N_a \mathcal{D}_t T^a=-\frac{V_N}{\dot{\phi}_0}, \label{T_eq}
\end{align}
where $V_T \equiv T^a V_a$ and $V_N\equiv N^a V_a$. From the second equation, the turn rate can be rewritten as
\begin{align}
\eta_{\perp}=\frac{V_N}{\dot{\phi}_0H}.   \label{turn_2}
\end{align}
Furthermore, as noticed in Ref.~\cite{Achucarro:2010da}, it can also be written in a suggestive way: 
\begin{align}
 \left|\eta_{\perp}\right|=\sqrt{2 \epsilon}\Mpl \kappa, \quad  \kappa\equiv\left(G_{ab} \frac{\mathcal{D} T^a}{d \phi_0} \frac{\mathcal{D} T^b}{d \phi_0}\right)^{1 / 2}.\label{turn_3}
\end{align}
Here $\kappa$ corresponds to the curvature of the trajectory which is velocity independent. Since the inflaton velocity $\sqrt{2\epsilon}\Mpl$ (measured by the e-fold) is constrained by the observation, one needs sub-Planckian scale $\kappa^{-1}$, in order to realize a large turn rate~\cite{Aoki:2024jha}. In addition, taking time derivatives of $T^aT_a=N^aN_a=1$ and $T^aN_a=0$, we obtain the evolution equations for $T^a$ and $N^a$:
\begin{align}
\mathcal{D}_t T^a=-H \eta_{\perp} N^a, \quad \mathcal{D}_t N^a=H \eta_{\perp} T^a.    \label{DtTN}
\end{align}

The slow-roll parameters are defined as
\begin{align}
\epsilon=-\frac{\dot{H}}{H^2}=\frac{\dot{\phi}_0^2}{2\Mpl^2H^2}, 
\qquad \eta^a=-\frac{\mathcal{D}_t \dot{\phi}_0^a}{H \dot{\phi}_0},  \label{epsilon} 
\end{align}
where we used Eq.~\eqref{EOM3}. The vector $\eta^a$ can be decomposed along the $T$ and $N$ directions:
\begin{align}
\eta^a=\eta_{\|} T^a+\eta_{\perp} N^a,    
\end{align}
with
\begin{align}
\eta_{\|}=-\frac{\ddot{\phi}_0}{H \dot{\phi}_0}, \qquad \eta_{\perp}=\frac{V_N}{\dot{\phi}_0 H}.    
\end{align}
The slow-roll condition requires $\epsilon, \eta_{\|}\ll 1$, while $\eta_{\perp}$ need not be small. In the single-field limit, $\eta_{\|}$ is related to the usual slow-roll parameter $\eta$~\eqref{def_SR} as
\begin{align}
\eta\equiv \frac{\dot{\epsilon}}{H\epsilon}= -2\eta_{\|}+2\epsilon.   \label{eta}
\end{align}

\subsection{Perturbation at quadratic order}
Here we discuss fluctuations around the background. 

For the metric, we apply the Arnowitt-Deser-Misner (ADM) decomposition~\cite{Arnowitt:1962hi},
\begin{align}
\mathrm{d} s^2=-N^2 \mathrm{d} t^2+h_{i j}\left(\mathrm{d} x^i+N^i \mathrm{d} t\right)\left(\mathrm{d} x^j+N^j \mathrm{d} t\right),   
\end{align}
with $N, N^i$, and  $h_{ij}$ being lapse, shift, and induced three-dimensional metric on constant-$t$ hypersurface, respectively. Then, we expand them around the FLRW background by
\begin{align}
N=1+\alpha(t,x), \quad N_i=\partial_i \beta(t,x), \quad h_{i j}=a^2(t)e^{2\zeta(t,x)} \delta_{i j},   \label{m_e} 
\end{align}
where $\alpha, \beta$ and $\zeta$ denote the small (scalar-) perturbations around the FLRW. Here we omitted the vector- and tensor-perturbations. 

For the field expansion, we expand the fields around the background as\footnote{Here we adopt the naive expansion as in Eq.~\eqref{f_e}, while the covariant expansion~\cite{Gong:2011uw, Elliston:2012ab} is useful for discussing the action for fluctuations at higher order.}
\begin{align}
\phi^a=\phi_0^a(t)+\varphi^a(t,x),\label{f_e}
\end{align}
where $\varphi^a$ correspond to scalar fluctuations. Furthermore, we decompose $\varphi^a$ by tangent $T$ and normal direction $N$,
\begin{align}
\varphi^a=\varphi T^a+\sigma N^a, \label{dec}   
\end{align}
where $\varphi$ and $\sigma$ denote the adiabatic perturbation and the isocurvature perturbation, respectively. 

The expanded metric~\eqref{m_e} and fields~\eqref{f_e} contain a redundancy, as one scalar degree of freedom can be eliminated by a coordinate transformation. Two well-known gauge choices are the comoving gauge $\varphi=0$ and the spatially flat gauge $\zeta=0$. Both approaches lead to the same result~\eqref{S_quad_a}, as expected.

\subsubsection*{Spatially flat gauge: $\zeta=0$}
By substituting the metric~\eqref{m_e} and field expansion~\eqref{f_e} into the action~\eqref{L_2}, we obtain the perturbed action for $\alpha,\beta,$ and $\varphi^a$. It turns out that the metric perturbations $\alpha$ and $\beta$ appear as auxiliary field without time derivative, which can be integrated out~\cite{Maldacena:2002vr},
\begin{align}
\alpha=\frac{H}{\dot{\phi}_0} \epsilon T^a\varphi_a, \quad \frac{\partial_i^2 \beta}{a^2}=-\frac{1}{2 H M_P^2}\left[\left(6 M_P^2 H^2-\dot{\phi}_0^2\right) \alpha+\dot{\phi}_0 T_a \mathcal{D}_t \varphi^a+V_a \varphi^a\right].  \label{ab}  
\end{align}
Plugging them back to the action, the quadratic action described only by physical quantities $\varphi^a$ is given by~\cite{Langlois:2008mn}
\begin{align}
S^{(2)}=\int d^4 x a^3\left[\frac{1}{2} G_{a b} \mathcal{D}_t \varphi^a \mathcal{D}_t \varphi^b-\frac{1}{2 a^2} G_{a b} \partial_i \varphi^a \partial_i \varphi^b-\frac{1}{2} M_{a b}^2 \varphi^a \varphi^b\right],    \label{qa}
\end{align}
where the mass matrix is given by
\begin{align}
M_{a b}^2=\nabla_b V_a-\dot{\phi}_0^2 T^c T^d \mathcal{R}_{a c d b}+\frac{2 H}{\dot{\phi}_0} \epsilon\left(V_a T_b+V_b T_a\right)+2(3-\epsilon) \epsilon H^2 T_a T_b,    
\end{align}
where $\mathcal{R}_{abcd}$ is the field-space Riemann curvature tensor. Note the second terms is reduced to $-\dot{\phi}_0^2 T^c T^d \mathcal{R}_{a c d b}=-\mathcal{R}\dot{\phi}_0^2N_aN_b/2$ for two-field case. 

Inserting the decomposition~\eqref{dec} into Eq.~\eqref{qa}, we arrive at
\begin{align}
S^{(2)}=\int d^4 x a^3&\left[\frac{\dot{\varphi}^2}{2}-\frac{\left(\partial_i \varphi\right)^2}{2 a^2}-\frac{m_{\varphi}^2}{2} \varphi^2+\frac{\dot{\sigma}^2}{2}-\frac{\left(\partial_i \sigma\right)^2}{2 a^2}-\frac{m_\sigma^2}{2} \sigma^2+\mathscr{L}_{\text {mix }}^{(2)}\right],    \label{quadaction}
\end{align}
where
\begin{align}
&m_{\varphi}^2=V_{T T}+2 \epsilon\left(3-\epsilon\right) H^2+4\epsilon HV_T/\dot{\phi}_0- \eta_\perp^2 H^2,\\
&m_\sigma^2=V_{N N}-\eta_\perp^2H^2 +\epsilon H^2 \Mpl^2\mathcal{R},\label{m_sigma}
\end{align} 
and 
\begin{align}
\mathscr{L}_{\text {mix }}^{(2)}   = \eta_\perp H\left(2 \sigma \dot{\varphi}-\eta H \sigma \varphi\right). \label{L_2_mix}
\end{align}
This agrees with the results in Ref.~\cite{Gong:2016qmq}. Note that we do not use any slow-roll approximation for the derivation. The action~\eqref{S_quad_a} for curvature perturbation $\zeta$ and isocurvature $\sigma$ can be obtained by replacing
\begin{align}
 \varphi=-\frac{H}{\dot{\phi}_0}\zeta.   
\end{align}

\subsubsection*{Comoving gauge: $\varphi=0$}
This approach provides a more straightforward derivation of Eq.~\eqref{S_quad_a} (see Ref.~\cite{Garcia-Saenz:2019njm} for further details).  
The procedure is the same: insert the fluctuations~\eqref{m_e} and~\eqref{f_e} with $\varphi=0$, then solve the algebraic equations for the lapse $\alpha$ and shift $\beta$, yielding
\begin{align}
\alpha=\frac{\dot{\zeta}}{H}, \quad 
\beta=-\frac{\zeta}{H}+\chi, \quad 
{\rm with}\quad \frac{\partial_i^2 \chi}{a^2}=\epsilon \dot{\zeta}- \frac{\eta_\perp\dot{\phi}_0}{\Mpl^2} \sigma.    
\end{align}
Substituting these back into the expanded action, we recover Eq.~\eqref{S_quad_a}.

\section{Explicit Forms of $n_s$ and $\alpha_s$}
\label{app:full}

Here we present the full expressions for the spectral index $n_s$ and the running of the spectral index $\alpha_s$ in the presence of a heavy field with mass $\tilde{m} = m/H$ and a mixing coupling $\tilde{\rho} = \rho/H$. 
As discussed in the main text, the effects of the heavy field appear through the sound speed $c_s(\tilde{m}, \tilde{\rho})$.

\subsection{Correction to spectral index $n_s$}
The corrections to spectral index $n_s$ are given by (see Eq.~\eqref{n_s}) 
\begin{align}
n_s=1-2 \epsilon-\eta+\frac{\partial \ln c_s^{-1}}{\partial \ln \tilde{\rho}} \frac{\eta}{2} +\frac{\partial \ln c_s^{-1}}{\partial \ln \tilde{m}} \epsilon,
\end{align}
where  
\begin{align}
&\frac{\partial \ln c_s^{-1}}{\partial \ln \tilde{\rho}}=\frac{2\mathcal{C}^{4}\tilde{\rho}^{2}}{\mathcal{A}\mathcal{D}},\\
&\frac{\partial \ln c_s^{-1}}{\partial \ln \tilde{m}}=-
\frac{\mathcal{C}^{4}\,\tilde{m}^{2}}{\mathcal{A}\mathcal{D}}
\left(
\mathcal{A}-\mathcal{C}^{4}(\tilde{m}^{2}-2)-1
\right).
\end{align}
Here $\mathcal{C}=16\pi/(\Gamma[-1/4])\sim 2.09$ is a number and we also defined \begin{align}
\mathcal{A}\equiv \sqrt{\mathcal{B}+4\mathcal{C}^{4}\tilde{\rho}^{2}},\quad
\mathcal{B}=\left(\mathcal{C}^4\left(\tilde{m}^2-2\right)+1\right)^2,\quad
\mathcal{D}\equiv \mathcal{A}-\mathcal{C}^4\left(\tilde{m}^2-2\right)+1.
\end{align}

\subsection{Correction to running of spectral index $\alpha_s$}
The corrections to the running of spectral index $\alpha_s$ are given by (see Eq.~\eqref{a_s})
\begin{align}
\nonumber \alpha_s =&\ -2\epsilon \eta -\eta \eta_2+\frac{\partial^2 \ln c_s^{-1}}{\partial^2 \ln \tilde{\rho}}  \frac{\eta}{2}+\frac{\partial^2 \ln c_s^{-1}}{\partial \ln \tilde{\rho}\ \partial \ln \tilde{m}} \left(\epsilon+\frac{\eta}{2}\right)+\frac{\partial^2 \ln c_s^{-1}}{\partial^2 \ln \tilde{m}}\epsilon\\
&+ \frac{\partial \ln c_s^{-1}}{\partial \ln \tilde{\rho}} \frac{\eta\eta_2}{2}+\frac{\partial \ln c_s^{-1}}{\partial \ln \tilde{m}} \epsilon \eta,
\end{align}
where 
\begin{align}
&\frac{\partial^2 \ln c_s^{-1}}{\partial^2 \ln \tilde{\rho}}=
\frac{4\mathcal{C}^{4}\tilde{\rho}^{2}}{\mathcal{A}^{3} \mathcal{D}^{2}}
\left(
- 2 \mathcal{C}^8\left(\tilde{m}^2-2\right) \tilde{\rho}^2
+ \mathcal{B} \mathcal{A}
- \mathcal{C}^{4}(\tilde{m}^{2}-2) \mathcal{B}
+ \mathcal{B}
+ 2 \mathcal{C}^{4} \tilde{\rho}^{2}
\right)
,\\
&\frac{\partial^2 \ln c_s^{-1}}{\partial \ln \tilde{\rho}\ \partial \ln \tilde{m}}=
\frac{8 \mathcal{C}^{8} \tilde{m}^{2} \tilde{\rho}^{2}}{\mathcal{B}^{3/2} \mathcal{D}^{2}} 
\left(
\mathcal{C}^{8} (\tilde{m}^{2}-2)^{2}
+ 
2 \mathcal{C}^{4}(\mathcal{A} + \tilde{\rho}^{2} - 1)
- \mathcal{C}^{4}\tilde{m}^{2} (\mathcal{A}-1)
- \mathcal{A}
\right)
,\\
\nonumber &\frac{\partial^2 \ln c_s^{-1}}{\partial^2 \ln \tilde{m}}=- 
\frac{8 \mathcal{C}^{8} \tilde{m}^{2}}{\mathcal{B}^{3/2} \mathcal{D}^{2}}
\Big(
\mathcal{A} 
-\mathcal{C}^{12}(\tilde{m}^{2}-2)^3
+ 
\mathcal{C}^{8} \tilde{m}^{2} (\tilde{m}^{2}-2) (\mathcal{A}+2\tilde{\rho}^{2}-3) \\
\nonumber   &\ \ \ \ \ \ \ \ \ \ \ \ \ \ \  - 2 \mathcal{C}^{8} (\tilde{m}^{2}-2) (\mathcal{A}+3\tilde{\rho}^{2}-3)
- 4 \mathcal{C}^{4} \mathcal{A} + 4\mathcal{C}^{4}  \tilde{\rho}^{2} (\mathcal{A}+\tilde{\rho}^{2}-2)\\
&\ \ \ \ \ \ \ \ \ \ \ \ \ \ \  + \mathcal{C}^{4} \tilde{m}^{2} (-2 \tilde{\rho}^{2} \mathcal{A} + 2 \mathcal{A} + \tilde{\rho}^{2}-3) + 6\mathcal{C}^{4} 
+ \tilde{\rho}^{2}-1
\Big).
\end{align}

\end{appendix}

\bibliographystyle{JHEP}
\bibliography{Refs}

\end{document}